\documentclass[journal, twocolumn, 9pt]{IEEEtran}

\usepackage{cite}
\usepackage{url}
\usepackage{physics}
\usepackage{amsthm}
\usepackage{empheq}
\usepackage{subcaption}
\usepackage{listings}
\usepackage{tcolorbox}
\usepackage{amsfonts}
\newtheorem{thm}{Theorem}

\newtheorem{cor}{Corollary}
\newtheorem{rmk}{Remark}

\newcommand{\R}{{\mathbb R}}
\newcommand{\C}{{\mathbb C}}

\newcommand{\sv}{{{\hat s}}}
\newcommand{\rv}{{{\vec r}}}

\newcommand{\Pe}{P_{\text{e|s}}}
\newcommand{\Peb}{P_{\text{\rm e, bin}}}
\newcommand{\Pbin}{P_{\mathrm{e, bin}}^{m'\vert m}}

\renewcommand{\d}{N}  
\newcommand{\nvar}{\sigma^2}

\newcommand{\Prob}{{P}}
\newcommand{\Pden}{{\mathcal P}}

\DeclareMathOperator*{\argmax}{argmax}

\newcommand{\dD}{{\tt d}_{\text{ic}}}
\newcommand{\dC}{{\tt d}_{\text{\rm c}}}
\newcommand{\dHS}{{\tt d}_{\text{HS}}}
\newcommand{\dStok}{{\tt d}_{\text{Stokes}}}

\newcommand{\thetaD}{\theta_{\text{\rm ic}}}
\newcommand{\thetaC}{\theta_{\text{\rm c}}}

\renewcommand{\aa}{{\tt a}}
\newcommand{\bb}{{\tt b}}
\DeclareMathOperator{\erfc}{erfc}

\newcommand{\Ce}{{\tt e}}
\newcommand{\Cf}{{\tt f}}
\newcommand{\Bc}{{\tt g}}

\newcommand\addtag{\refstepcounter{equation}\tag{\theequation}}

\usepackage[noprefix]{nomencl}
\makenomenclature

\begin{document}
%
\title{Mode Vector Modulation: Extending Stokes Vector Modulation to Higher Dimensions}
%
%
%

\author{Jaroslaw~Kwapisz, 
        Ioannis~Roudas,
        and~Eric~Fink
\thanks{I. Roudas is with the Department of Electrical and Computer Engineering, Montana State University, Bozeman, MT, 59717 USA, ioannis.roudas@montana.edu.}
\thanks{J. Kwapisz and E. Fink are with the Department of Mathematical Sciences, Montana State University.}
}

\maketitle

\begin{abstract}
\footnotesize
 The use of multidimensional modulations can 
 decrease the energy consumption of optical links. In this paper, we propose and study Mode Vector Modulation (MVM), a generalized polarization modulation scheme for transmission over multimode/multicore optical fibers or free space. Similar to Polarization Shift Keying (PolSK) and Stokes Vector Modulation (SVM), MVM can be used in conjunction with direct detection and, therefore,  is suitable for next-generation, short-haul optical interconnects.
This paper focuses on the MVM transceiver architecture, the back-to-back performance of optically-preamplified MVM direct-detection (DD) receivers, the optimized geometric shaping of the MVM constellation, and the related bit-to-symbol mapping.
We show that MVM DD outperforms conventional single-mode, direct-detection-compliant, digital modulation formats by several dB's in terms of receiver sensitivity and the SNR gain increases with the number of spatial degrees of freedom (SDOFs) $N$. 
\end{abstract}

\section{Introduction}
Over the last few years, there has been a keen interest in developing inter-data-center optical interconnects with bit rates approaching or exceeding 1 Tb/s 
\cite{Eiselt18,Perin2018,PerinJLT21,hu_ultrahigh-net-bitrate_2022}. 
Low cost, low energy consumption, and high spectral efficiency compared to binary intensity modulation/direct-detection are prerequisites for Tb/s short-haul applications \cite{Eiselt18}.  To satisfy the first two specifications, 
direct-detection optical receivers are currently preferred 
to coherent optical receivers. 
To increase spectral efficiency, binary intensity modulation (IM) is gradually replaced by $M$-ary pulse amplitude modulation ($M$-PAM) \cite{Pang:20, Le:JLT:2020, wettlin_dsp_2020, ozolins_100_2021}. 

Given the forecasted exponential increase in data traffic in the near future due to broadband applications \cite{Cisco20}, to accommodate traffic demands, it will be important to keep increasing  the spectral efficiency per fiber lane or per wavelength channel of short-haul optical links in an energy-efficient manner. The main disadvantage of $M$-PAM is that its energy consumption scales quadratically with the number of amplitude levels $M$ \cite{Proakis},  since the $M$-PAM constellation is one-dimensional. Therefore, it is imperative to adopt higher-dimensional modulation formats other than PAM, which are still amenable to direct detection, to keep upfront cost down.

Therefore, looking forward, we anticipate that it will be necessary to modulate additional attributes  of the optical wave other than the amplitude, e.g., the phase or the polarization, in order to increase spectral efficiency beyond today's values. Consequently, it will be necessary to recover the information imprinted in the electric field of the optical wave using self-homodyning \cite{mecozzi2016kramers, chen2020high, yoshida_JLT_19, Chen_JLT_20, ji_theoretical_2021}. 
This is predicated on the assumption that the price of phase- and polarization-diversity coherent optical receivers will continue to be prohibitive for short-haul applications in the medium- to long-term, which is currently subject of debate  \cite{Perin2018, PerinJLT21, zhou_beyond_2020, morsy-osman_dsp-free_2018, jia_coherent_2021, rizzelli_scaling_2021}. Based on the above considerations, spectrally-efficient modulation formats for short-haul optical communications systems with higher dimensionality than $M$-PAM 
\cite{plant2018trends}, in conjunction  with multi-branch, self-homodyne, direct-detection receivers \cite{chen2020high,yoshida_JLT_19,Chen_JLT_20}, have become one of the most active research areas in contemporary optical communications. 

Let us discuss self-homodyning in direct-detection-based interferometry and  direct-detection-based polarimetry. Direct-detection-based interferometric receivers employ delay interferometers, like the ones used in differential receivers  for Differential Phase Shift Keying (DPSK) and Differential Quadrature Phase-Shift Keying (DQPSK) \cite{Ho2005}. Direct-detection polarimeters, on the other hand, are based on optical components performing polarization transformations (e.g., polarization beam splitters, polarization rotators) \cite{Huard1997} but the final result is also interferometric in nature, as explained  below. 

\begin{figure*}[!ht]
	\centering
		\includegraphics[width=1.0\textwidth]{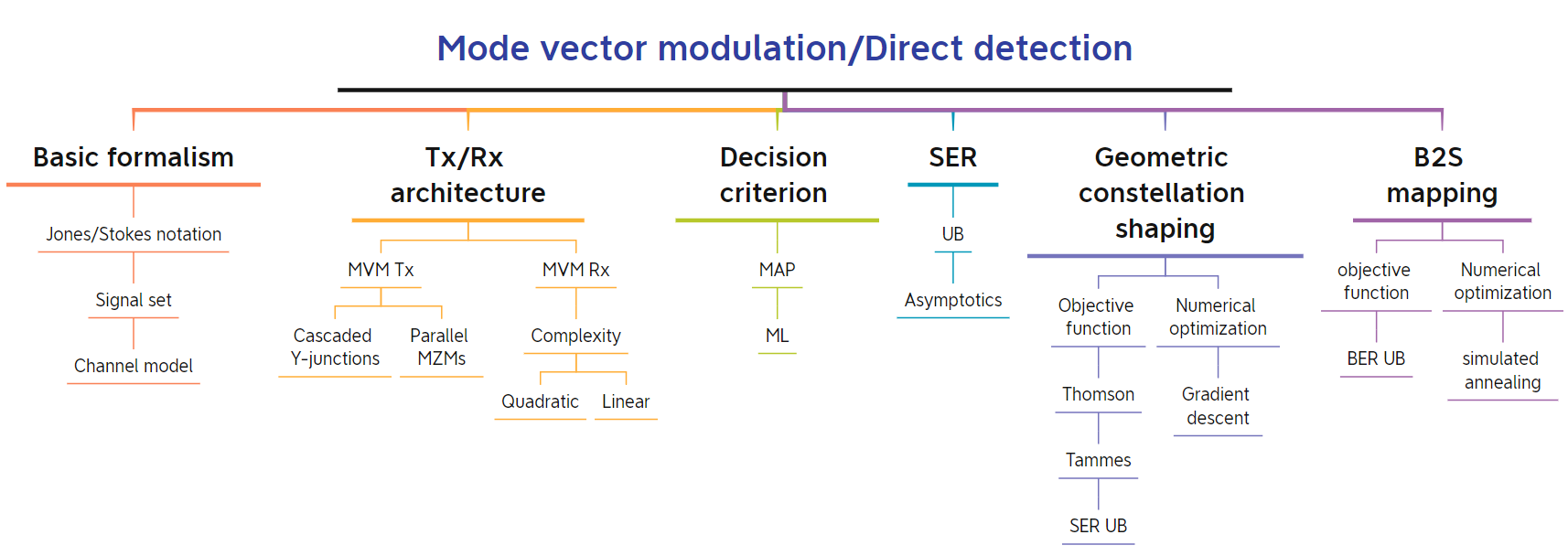}
	\caption{Hierarchical organization of the paper (Abbreviations: MVM=Mode vector modulation, Tx=transmitter, Rx=receiver, UB=Union Bound, BER=bit error rate, SER=symbol error rate, MAP= maximum a posteriori, ML=maximum likelihood, MZM=Mach-Zehnder modulator, B2S mapping=bit-to-symbol mapping).}
	\label{fig:VisualAbstract}
\end{figure*}

In both direct-detection-based interferometry and  direct-detection-based polarimetry, the optical wave reaching the optical receiver is initially divided into multiple copies, which are subsequently modified by a series of optical devices, and then added pairwise before detection. The last step is crucial: Since photodiodes can be modeled as square-law detectors, the intensities of these sums of terms contain beating products which {depend on} 
 the phase of the optical wave. In fact, these beating terms are mathematically very similar to the ones provided  by coherent homodyne receivers. In other words,  under certain conditions, interferometric  and polarimetric direct-detection receivers enable access to all the attributes of the optical wave {apart from the overall phase} at relatively low cost and complexity compared to their coherent optical counterparts.



Let us turn our attention to digital polarization modulation formats. 
Polarization shift keying (PolSK) was first studied in the late 1980s  \cite{betti1990multilevel, betti1992polarization} and early 1990s \cite{Benedetto_1992, benedetto1994multilevel, Benedetto_1995} before later falling into obscurity. It was recently revived as a subset of Stokes vector modulation (SVM) \cite{Kikuchi2014, Cercos2015a, Che2015, Che2016a, Dong2016, MorsyOsman2016, Chagnon2016, Chagnon2017, Sowailem2017, Ishimura2018, Ghosh2018, Hoang2018, TANEMURA2018, MorsyOsman2019, Ghosh2019, Ishimura2019, Cui2019, Kikuchi2020, Tanemura2020, Huo2020, Jin2020} when research in direct-detection systems was rekindled. This renewed interest in digital polarization modulation formats 
has been fueled by the maturity and low-cost of integrated photonics components and the possibility of using adaptive electronic equalizers in the  direct-detection optical receivers to compensate  for polarization rotations introduced by short optical fibers\footnote{Other applications of the Stokes space formalism were also proposed in combination with various modulation formats  \cite{Nazarathy_2006, Nazarathy2006, Nazarathy2007, perrone2018multidimensional, Ziaie2019, Ji2019, Feuer2020}, as well as  in combination with digital signal processing (DSP) \cite{Visintin2014, Caballero2017, Fernandes_2017, Che2019, Eltaieb2020}.}.


SVM allows for more power-efficient signaling than $M$-PAM. This is achieved by spreading the constellation points in the three-dimensional Stokes space, as opposed to the one-dimensional $M$-PAM  signal space. 

To further increase the energy-efficiency of SVM formats, a transition to a higher-dimensional Stokes space is necessary, which can be achieved by using SVM in conjunction with few-mode and multicore fibers or free space \cite{Ji2019}. We call this format mode vector modulation (MVM).

SVM and MVM are spatial modulation formats \cite{Wen2019}. It is worth mentioning here that several papers studied spatial modulation formats for {\em coherent} optical communication systems over MCFs. For instance, Eriksson et al. \cite{eriksson2014k} analyzed multidimensional position modulation (MDPM) with multiple pulses per frame ($K$-over-$L$-MDPM) in combination with quadrature phase shift keying (QPSK), polarization multiplexed QPSK (PM-QPSK) and polarization-switched QPSK (PS-QPSK)  to increase both the spectral efficiency and the asymptotic power efficiency compared to conventional modulation formats. In companion  papers, Puttnam et al. \cite{Puttnam2014, puttnam2014energy, Puttnam2017} reviewed spatial modulation formats for high-capacity coherent optical systems using homogeneous multicore fibers.

In this paper, we study, for the first time, short-haul,  optical interconnects using MVM, in conjunction with optically-preamplified  direct-detection receivers. A  visual abstract of the paper is given in Fig. \ref{fig:VisualAbstract}. In the remainder of the paper, we elaborate on the following topics: 
\begin{enumerate}
    \item An overview of MVM along with the necessary mathematical formalism, notation, and simplifying assumptions (Sec. \ref{sec:MVMoverview});
    \item The optimal MVM transceiver architecture (Sec. \ref{sec:txrx_design});
    \item The performance limits of MVM optically-preamplified, direct-detection receivers using both Monte Carlo simulation and a new analytical formula that we derived for the union bound (Sec. \ref{sec:Pes});
    \item The  design of geometrically-shaped constellations with arbitrary cardinality $M$, obtained by numerical optimization of  various objective functions using the method of gradient descent (Sec. \ref{sec:GeometricShaping});
    \item The bit-to-symbol mapping optimization using simulated annealing (Sec. \ref{sec:b2s});
    \item The investigation of various constellation designs and bit encodings using analytical and numerical methods (Sec. \ref{sec:results});
    \item The use of simplex MVM constellations based on symmetric, informationally complete, positive operator-valued measure (SIC-POVM) vectors \cite{Fuchs} (Sec. \ref{sec:results}).
\end{enumerate}
Early results were presented in \cite{Roudas:ECOC21, Fink:IPC21, Kwapisz:CLEO22, Kwapisz:IPC22}.

\section{Mode Vector Modulation Overview}\label{sec:MVMoverview}
\subsection{MVM signal representation}
\noindent  MVM  can be used together with multimode  and multicore fibers, as well as for free-space transmission. In this section, for the description of the operation of the  MVM transceiver, without loss of generality, we examine the special case of MVM transmission over an ideal homogeneous  multicore fiber with identical single-mode cores and negligible differential  group delay among cores.

\begin{figure}[!htb]
	\centering
	\includegraphics[width=3in]{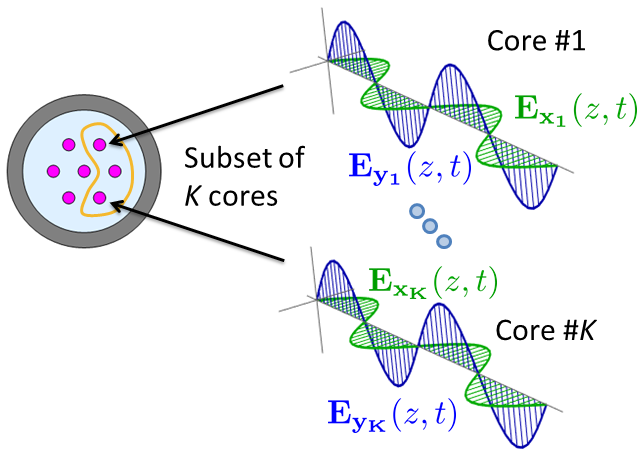}
	\caption{MVM over homogeneous MCFs with single-mode cores.}
	\label{fig:MVM_MCFs}
\end{figure}

\begin{figure}[!htb]\centering
\captionsetup[subfigure]{justification=centering}
    \begin{subfigure}[b]{.45\columnwidth}
    \centering
	    \includegraphics[width=1.5in]{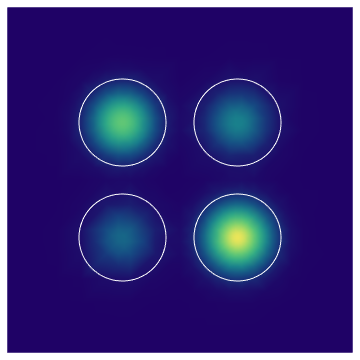}
	    \caption{}
	\end{subfigure}
	\begin{subfigure}[b]{.45\columnwidth}
	    \centering
	    \includegraphics[width=1.5in]{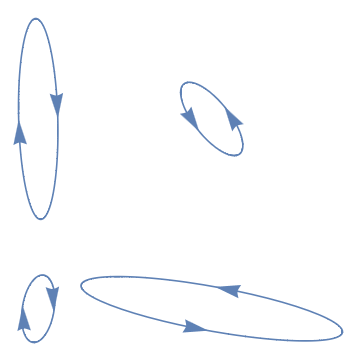}
	    \caption{}
	\end{subfigure}
	\caption{(a)  Intensity plot and (b) Polarization ellipses  of an MVM signal propagating over an ideal four-core MCF with identical uncoupled single-mode cores.} 
	\label{fig:intensityPlot_M64_N8.png}
\end{figure}

We assume that we select a subset of $K$  single-mode cores of the multicore fiber (Fig.~\ref{fig:MVM_MCFs}).
MVM modulation consists in sending optical pulses  over all these cores simultaneously with the same shape but different amplitudes and initial phases (Fig.~\ref{fig:intensityPlot_M64_N8.png}). Similar to SVM over single-mode fibers, wherein the optical wave can be analyzed in two orthogonal states of polarization, e.g., $x$ and $y$, the composite optical wave of MVM over a homogeneous  single-mode-core multicore fiber can be described by $N=2K$ orthogonal states of polarization, e.g., $x$ and $y$ in each core.

The mathematical representation of the MVM signals at the  fiber input  is written as
\begin{equation} 
	{\bf E}_{m} (t)=A_{m} \exp\left( \iota \phi _{m} \right) g(t) \ket{ s_m},
	\label{eq:MVMsigs} 
\end{equation}
where $m=1,\ldots,M$, $A_m$ and $\phi_{m}$ denote the common amplitude  and phase, respectively, $g(t)$ is a real function describing the pulse shape, and $\ket{ s_m } $ is a generalized unit Jones vector collecting the complex excitations of the cores, i.e., the amplitudes and phases of electric fields of the optical waves \cite{Antonelli:12,Roudas_PJ_17,Roudas_JLT_18}. 
    \nomenclature[$emt$]{$\mathbf{E}_m (t)$}{Electric field at fiber input}
    \nomenclature[$eg$]{$\mathcal{E}_g$}{Pulse energy}
	\nomenclature[$am$]{$A_m$}{Common signal amplitude}
	\nomenclature[$phim$]{$\phi_m$}{Common signal phase}
	\nomenclature[$gt$]{$g(t)$}{Pulse shape}
	\nomenclature[$sket$]{$\ket{s}$}{Generalized unit Jones vector} 

The signal energy $\mathcal{E}_s$ is given by \cite{Proakis} \nomenclature[$es$]{$\mathcal{E}_s$}{Signal energy}
\begin{align}
\mathcal{E}_s :=& \frac{1}{2}\int_{-\infty}^\infty {\bf E}_{m} (t)^\dagger {\bf E}_{m} (t) dt 
= \frac{A_m^2}{2}\int_{-\infty}^\infty g(t)^2dt= \frac{A_m^2}{2}\mathcal{E}_g , 
    \label{eq:sigEnergy}
\end{align}
where $\dagger$ denotes the adjoint (i.e., conjugate transpose) of a matrix and  $\mathcal{E}_g$ denotes the pulse energy defined as
\begin{equation}
\mathcal{E}_g := \int_{-\infty}^\infty g(t)^2dt.
    \label{eq:pulseEnergy} 
\end{equation}

In the remainder of the article, without loss of generality,  we consider that the common amplitude $A_m$ and phase $\phi_{m}$ in \eqref{eq:MVMsigs} are constant. In other words, we focus exclusively on a special case of MVM that is a generalization of PolSK to higher dimensions.

\subsection{Mathematical notation}
\noindent  Throughout the paper, we follow the conventions of \cite{Gordon2000, Roudas_PJ_17, Roudas_JLT_18}, where Dirac's bra-ket vectors represent both unit and non-unit vectors in the generalized Jones space, while hats indicate unit vectors and arrows indicate non-unit vectors in the generalized Stokes space. 

We can parameterize a unit Jones vector
$\ket{s}$, up to phase, using $2N-2$ hyperspherical coordinates \cite{Roudas_JLT_18}, i.e., 
\begin{align}
	\ket{s} &:= 
	\left[
	\cos \phi_{1}, \ \sin \phi_{1}\cos\phi_{2}\, e^{\iota \theta _{1}},  \sin\phi_{1}\sin\phi_{2}\cos\phi_{3}\, e^{\iota \theta _{2}},\ldots ,  \right.  \notag\\
	&\quad \left. 
	\sin\phi_{1} \cdots \cos\phi_{N-1}\, e^{\iota \theta _{N-2}}, \
	\sin\phi_{1}\cdots \sin\phi_{N-1}\, e^{\iota \theta _{N-1}} 
	\right]^{T} 
	\label{eq:jv}
\end{align}
where the superscript $T$ indicates transposition.



Unit Jones vectors, up to phase, are often represented by generalized real unit Stokes vectors $\hat s$ in a higher-dimensional real vector space $\R^{\d^2-1}$. Generalized unit Stokes vectors are defined by the quadratic form \cite{Roudas_JLT_18}
\begin{equation} 
	\hat s : = C_N \mel{s}{\mathbf{\Lambda }}{s},
	\label{eq:StokesVectorsdf}
\end{equation} 
where $\mathbf{\Lambda}$ denotes the  generalized Gell-Mann spin vector  and $C_N$ denotes the normalization coefficient \cite{Roudas_JLT_18} 
\begin{equation}
C_N := \sqrt{ \frac{N}{ 2\left(N-1\right)}}.
\end{equation}
    \nomenclature[$shat$]{$\hat{s}$}{Generalized unit Stokes vector}
	\nomenclature[$lambda$]{$\mathbf{\Lambda}$}{Generalized Gell-Mann spin vector}
	\nomenclature[$n$]{$N$}{Number of spatial and polarization degrees of freedom}
	\nomenclature[$cn$]{$C_N$}{Normalization coefficient, $C_N:= \sqrt{N/[2(N-1)]}$}

From the generalized Stokes vector definition \eqref{eq:StokesVectorsdf}, we notice that the dimensionality of the generalized Stokes space grows quadratically with the number of spatial and polarization modes in MMFs/MCFs. Therefore, instead of using SVM-DD in conjunction with the conventional 3D Stokes space, we can generate more energy-efficient  constellations by spreading the constellation points in the generalized Stokes space.

Also notice that the $(N^2-1)$ components of $\hat s$ are functions of the $2N-2$ hyperspherical coordinates of $\ket{s}$  in \eqref{eq:jv} and, therefore,  are  interdependent.


For each Jones vector $\left.|s\right\rangle $, we can define the associated projection operator ${\bf S} := | s \rangle \langle s |$, which represents a mode filter, i.e., the equivalent of a polarizer in the two-dimensional case. 
This projection operator can be expressed in terms of the identity matrix and the generalized Gell-Mann matrices \cite{Roudas_JLT_18} 
\begin{equation}
\label{jonesToop:eq}
{\bf S} = \frac{1}{N}{\bf I }+\frac{1}{2 C_N} \hat{s}\cdot {\bf{\Lambda }}.
\end{equation}
    \nomenclature[$s$]{$\mathbf{S}$}{Projection operator dyad, $\mathbf{S}:=\ket{s}\bra{s}$}

By rearranging the terms in \eqref{jonesToop:eq}, we obtain
\begin{equation}
\label{StokesCoefficients:eq}
\hat{s}\cdot {\bf{\Lambda }}=2 C_N\left({\bf S} - \frac{1}{N}{\bf I }\right) .
\end{equation}
 From \eqref{StokesCoefficients:eq}, we see that Stokes vectors arise as  coefficients with respect to a fixed Gell-Mann basis for the \emph{trace neutralized dyad} ${\bf S}-\frac{1}{\d}{\bf I}_N$, assuming  $\ip{s}=1$. 
 
 In the remainder of the article, we will use the Jones vector up to phase $e^{\iota \theta}\ket{s}$, the dyad  ${\bf S}=\dyad{s}$, and the Stokes vector $\hat s$ interchangeably, depending on which one is more convenient.
  In particular, even when we use Jones vectors to represent points in a constellation, since we consider noncoherent detection, we refer to it as a generalized Stokes constellation.

\subsection{Simplifying assumptions\label{sec:assumptions}}

\noindent In Sec. \ref{sec:Pes}, we will analytically calculate  the back-to-back performance of $M$-ary MVM over $N$ spatial and polarization degrees of freedom in the amplified spontaneous emission (ASE) noise-limited regime. For mathematical tractability, we neglect all transmission impairments other than ASE noise and random carrier phase shifts, as well as transceiver imperfections and implementation penalties. These simplifying assumptions are justified in the sense that we want to  quantify the ultimate potential of MVM for use in optical interconnects. 

Nevertheless, it is worth discussing upfront about the anticipated impact of the most prominent transmission effects. 

In general, the extension of SVM to MVM requires similar conditions for transmission, i.e., negligible chromatic dispersion (CD), modal dispersion (MD), and mode-dependent loss (MDL), or their full compensation, either in the optical or the electronic domain, before making decisions on the received symbols at the receiver. Let us briefly contemplate how feasible it would be to satisfy these requirements in the case of practical homogeneous  multicore fibers with single-mode cores. 

As a starting point, consider transmission over homogeneous MCFs with uncoupled or weekly-coupled single-mode cores. These fibers typically exhibit static and dynamic intercore skew \cite{Puttnam2014}. The static differential mode group delay (DMGD) spread is on the order of 0.5 ns/km and grows linearly with the transmission distance. The DMGD spread due to the dynamic component of the intercore skew is of the order of 0.5 ps/km and also grows linearly with the transmission distance. 

On the other hand, coupled-core MCFs exhibit modal dispersion and strong coupling among their supermodes and the DMGD grows with the square root of the transmission distance \cite{Hayashi:17, Yoshida:19}. From  published values based on the characterization of several  coupled-core MCFs used in MDM experiments, we conclude that the MD coefficient is currently on the order of 3-6 ps/$\sqrt{\mathrm{km}}$. These values are much higher than typical PMD coefficient values for SMFs, e.g., from the data sheet of Corning\textregistered \, {SMF-28 Ultra}\textregistered \,  optical fiber \cite{SMF28u_datasheet}, we notice that the PMD coefficient  is less than 0.1 ps/$\sqrt{\mathrm{km}}$.

Transmission impairments can be compensated using a combination of optical and electronic techniques at the transmitter and the receiver. These techniques are out of the scope of this paper, since we are interested in the back-to-back performance of MVM systems, and will be part of future work. For simplicity, in the depiction of the optically-preamplified MVM direct-detection receiver in Fig.~\ref{TxRx:sub2}, we assume ideal optical post-compensation of all transmission impairments.

\section{Transceiver Design}\label{sec:txrx_design}
\begin{figure*}[ht!]
	\centering
	\captionsetup[subfigure]{justification=centering}
	\begin{subfigure}[t]{0.5\textwidth}
		\centering
		\includegraphics[width=.8\linewidth]{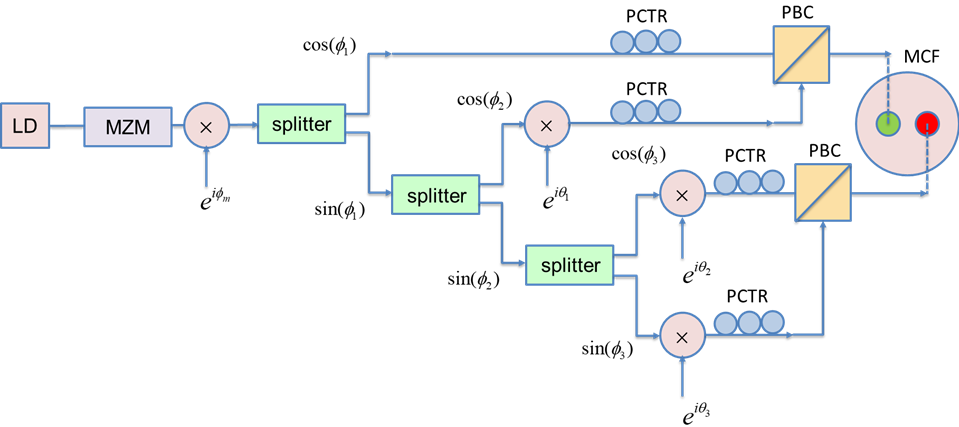}
		\caption{}
		\label{TxRx:sub1}
	\end{subfigure}%
	\begin{subfigure}[t]{0.5\textwidth}
		\centering
		\includegraphics[width=.8\linewidth]{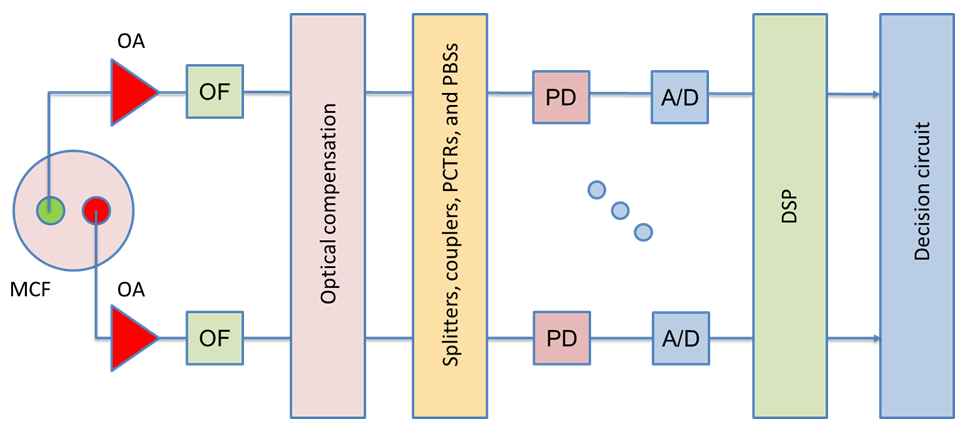}
		\caption{}
		\label{TxRx:sub2}
	\end{subfigure}
\vspace{0.1 cm}
	\caption{Schematic of the proposed (a) MVM transmitter and (b) Optically-preamplified MVM DD receiver. Symbols: LD=laser diode, MZM=Mach-Zehnder modulator, OA=optical amplifier, OF=optical filter,  PCTR=polarization controller, PBC/S=polarization beam combiner/splitter, PD=photodiode, A/D=Analog-to-digital converter. (Condition: $N=4$.)}
	\label{fig:TxRx}
\end{figure*}

 In Fig. \ref{TxRx:sub1}, we draw the block diagram of an MVM transmitter for an ideal homogeneous multicore fiber with two identical single-mode cores ($N=4$). The schematic shows the optical components required for a single wavelength, but the architecture can be easily generalized for wavelength division multiplexing (WDM). Our goal is to generate the $N$ spatial and polarization components of the MVM signal as described by \eqref{eq:MVMsigs}. 
 
 The transmitter design begins with a single semiconductor laser diode. Subsequently, a Mach-Zehnder modulator, followed by a phase modulator, can be employed to alter the pulse shape $g(t)$, as well as the common amplitude $A_m$ and phase  $\phi_m$ of the MVM signal according to  \eqref{eq:MVMsigs}. After that, electro-optic splitters  can be used to partition the signal into $N$ parallel branches. By adjusting the control voltage of each Y-junction, an arbitrary power splitting ratio between its two output ports can be achieved. Recalling the hyper-spherical parametrization of the unit Jones vector $\ket{s_m}$ in \eqref{eq:jv}, the power splitting ratio is $\cos^2 (\phi _{1} ):\sin^2 (\phi _{1} )$ at the first Y-junction, $\cos^2 (\phi _{2} ):\sin^2 (\phi _{2} )$ at the second Y-junction, and so forth\footnote{An alternative design, similar to the one proposed by Kikuchi and Kawakami for SVM \cite{Kikuchi:14}, would entail the use of a passive 1:$N$ splitter, followed by an array of $N$ parallel Mach-Zehnder modulators, one at each individual transmitter branch.}.  Then, an array of phase modulators is used to generate phase differences among vector components. Finally, polarization controllers and polarization beam combiners are used to merge pairs of signals originating from different optical paths to create orthogonal states of polarization (SOPs) that are launched into separate fiber cores.

Fig. \ref{TxRx:sub2} shows the direct-detection receiver for $N=4$. The purpose of the optical front-end is to measure the  Stokes components of the incoming spatial superchannel, which are given by \eqref{eq:StokesVectorsdf}. To begin, it is necessary to separate the spatial and polarization components of the individual tributaries of the spatial superchannel using mode demultiplexers and polarization beam splitters. Then, an array of $N$ photodiodes are used to measure their powers. In addition, polarization controllers and power splitters/couplers are used to combine the different spatial and polarization components pairwise in order to create $N(N-1)/2$ distinct combinations. The real and imaginary parts of the latter are measured using an array of $2N(N-1)$ identical photodiodes grouped in pairs. In total, $2N^2-N$ photodiodes are needed to measure all the elements of the dyad  ${\bf S}=\dyad{s}$ independently \cite{Kwapisz:IPC22}. 

However, by taking advantage of the interdependence of Stokes components, as they are functions of the $2N-2$ hyperspherical coordinates of $\ket{s}$, it is possible to reduce the direct-detection receiver front-end complexity. In \cite{Kwapisz:IPC22}, we showed that $O(N)$ photodiodes are sufficient to estimate the Stokes parameters of the spatial superchannel.

For the purposes of this article, we assume that the simplifying assumptions of Sec. \ref{sec:assumptions} hold, i.e., the optically-equalized communication channel exhibits negligible  CD, MD, and MDL  so that the residual transmission effects (random modal birefringence and random differential carrier phase shifts) can be modeled by a frequency-independent random unitary matrix. The  action of this transfer matrix is a dynamic rotation of the received mode vector that varies slowly over time. 

Stokes receiver DSP can estimate Stokes vector rotations caused by fiber propagation and can counteract them by multiplying, in Stokes space, the received generalized Stokes vector with a compensating generalized M\"uller matrix. Alternatively, it is possible to perform optical derotation and MD/PMD compensation of the received generalized Jones vector, which is driven by the Stokes vector receiver DSP.

The decisions of the Stokes vector receiver in Fig. \ref{TxRx:sub2} are based on the maximum a posteriori (MAP) criterion \cite{Proakis}, which is equivalent to the maximum-likelihood criterion for equiprobable signals \cite{Proakis} (see Sec. \ref{sec:decision_criterion}). Applying the maximum-likelihood criterion in the generalized Jones space, we will show that the optimum decision maximizes 
the modulus of the inner product of Jones vectors (cf. \eqref{eq:optScheme}).

\section{Symbol error Probability }\label{sec:Pes}
\subsection{Transmission channel model}
We formulate the transmission channel by a discrete-time model \cite{Divsalar_TCOM_90}. All optical and electronic signals from now on are represented by their samples taken once per symbol.

After optical post-compensation, we assume that all transmission impairments are fully compensated. For instance, let ${\bf U} $ denote the unitary {Jones transfer matrix} of the optical fiber due to modal birefringence. We  assume that the unitary fiber transfer matrix  ${\bf U}$ is fully compensated, up to a random phase $\theta$,  by a  zero-forcing optical adaptive equalizer with transfer matrix ${\bf W} $ so that
 \begin{equation}
  {\bf W U} = e^{\iota \theta} {\bf I}_N,
 \end{equation}
where ${\bf I}_N$ denotes the $N \times N$ identity Jones matrix.

After the optical front-end, at the $0-$th sampling instant, in the absence of noise, the incoherent receiver recovers  the transmitted Jones vector $\ket{s_m}$ up to phase $\theta$, which one could denote by $e^{\iota \theta}\ket{s_m}$.

  Optical amplifiers introduce amplified spontaneous emission (ASE) noise, which is modelled as additive white Gaussian noise (AWGN). 
  
  The received vector at the $0-$th sampling instant before photodetection equals \cite{Divsalar_TCOM_90} 
  \begin{equation}
  \label{eq:AWGNJoneschannelBis}
    \ket{r} = A_{m} e^{\iota \phi _{m} } e^{\iota \theta} \ket{s_m} + \ket{n}, 
 \end{equation}
\nomenclature[$rket$]{$\ket{r}$}{Received Jones vector, $\ket{r} = A_{m} e^{\iota \phi _{m} } e^{\iota \theta} \ket{s_m} + \ket{n}$}%
\nomenclature[$nket$]{$\ket{n}$}{Noise vector in Jones space}%
 where  $\theta$ is uniformly distributed over $[0, 2\pi)$ and $\ket{n}$ is a  noise vector in Jones space. Notice that $\ket{r},  \ket{n}$ are non-normalized Jones vectors, whereas $\ket{s_m}$ is a unit Jones vector.
 
 Assuming identical optical amplifiers at the output of all cores,  $\ket{n}$ has independent and identically distributed (i.i.d.) entries following a complex Gaussian distribution. The probability density function (pdf) of $\ket{n}$ is
 \begin{equation}
 \label{gaussianDistro:eq}
\Pden(\ket{n})=\frac{1}{(2 \pi \nvar)^\d} \exp\left(-\frac{\ip{n}}{2\nvar}\right),
 \end{equation} 
where $\nvar$ denotes the noise variance per quadrature after the matched optical filter. 

The noise energy in each spatial degree of freedom is
\begin{equation}
  \mathcal{N}_0=  \frac{1}{2}\int_{-\infty}^\infty {\bf n}_{\nu} (t)^\dagger {\bf n}_{\nu} (t) dt =\nvar 
\end{equation}
 where ${\bf n}_{\nu} (t)$ is the complex noise component in a single quadrature plane of the electric field (given by any $\ket{\nu} \in \C^N$). 

The symbol SNR $\gamma_s$, taking into account the noise over a spatial degree of freedom of the signal $\ket{s_m}$, is defined as
 \begin{equation}
  \gamma_s := \frac{\mathcal{E}_s}{\mathcal{N}_0}=\frac{A_m^2}{2\nvar}\mathcal{E}_g.
 \end{equation}
 In the following, without any loss of generality, we assume that $A_m=1$ and $\mathcal{E}_g=1$, so 
  \begin{equation}\label{gammasSigma:eq}
  \gamma_s=\frac{1}{2\nvar}.
 \end{equation}

The received dyad ${\bf R}$ is related to the transmitted dyad ${\bf S}_m=\dyad{s_m}$ by the \emph{Stokes channel} formula:
 \begin{equation}
  \label{eq:AWGNStokesChannelIncoherent}
   \underset{{\bf R}}{\underbrace{\dyad{r}}} = \underset{{\bf S}_m}{\underbrace{\dyad{s_m}}} +   \underset{{\bf N}_m}{\underbrace{2\Re\left(\ketbra{s_m}{n} \right) + \dyad{n} }}.
 \end{equation}
 Observe that, in this formulation, the last two terms form a non-Gaussian 
  noise ${\bf N}_m$ exhibiting   signal-noise and noise-noise beating 
 \begin{equation}
   \label{eq:StokesProjNoise}
   {\bf N}_m := 2\Re\left( \ketbra{s_m}{n} \right) + \dyad{n}.
   \end{equation}

   We  note  that Jones vectors, up to phase, are often represented by generalized real Stokes vectors $\sv$ in a higher-dimensional real vector space $\R^{\d^2-1}$. The latter arise as  coefficients with respect to a fixed Gell-Mann basis for \emph{trace neutralized dyad} ${\bf S}$, specifically, ${\bf S}-\frac{1}{\d}{\bf I}_N$ assuming normalization $\ip{s}=1$. We will use the Jones vector up to phase $e^{\iota \theta}\ket{r}$, the dyad  ${\bf R}=\dyad{r}$, and the Stokes vector $\rv$ interchangeably, depending on which one is more convenient.
   In particular, even when we use Jones vectors to express points of a constellation, we refer to it as a (generalized)  Stokes constellation when it is deployed in the incoherently received communication.


\subsection{Optimum decision criterion for equipower signals\label{sec:decision_criterion}}

\noindent The decision scheme  at the receiver uses the \emph{maximum a posteriori  probability} (MAP) criterion  \cite{Proakis} to select a signal $\sv_{\hat{m}}$ out of the set of $M$ transmitted signals given that $\rv$ was received
\begin{align}
  \hat{m} &:= \argmax_{1 \leq m \leq M} \Pden(\sv_m \ | \ \rv).
  \label{eq:MAP_criterion}
\end{align}

From Bayes' theorem, the conditional probability distribution  of $\sv_m$ given $\rv$ is 
\begin{equation}
  \Pden(\sv_m \ | \ \rv) = \frac{P_m \Pden(\rv \ | \ \sv_m)}{\Pden(\rv)},
  \label{eq:Bayes}
\end{equation}
where $P_m$ is the probability of sending $\sv_m$ ,  $\Pden(\rv)$ is the marginal pdf of receiving $\rv$, and $\Pden(\rv \ | \ \sv_m)$ is the  likelihood pdf of receiving $\rv$ given that $\sv_m$ was sent.



By substituting \eqref{eq:Bayes} into \eqref{eq:MAP_criterion} and omitting the common denominator $\Pden(\rv)$ (which does not influence the decision) we obtain
\begin{align}
  \hat{m}   
  &= \argmax_{1 \leq m \leq M} {P_m \Pden(\rv \ | \ \sv_m)}.
  \label{eq:MAP_criterion_2}
\end{align}

In this paper, we focus exclusively on equiprobable symbols, and, therefore, $P_m={1}/{M}.$ In this case, the   \emph{maximum a posteriori  probability} (MAP) criterion of \eqref{eq:MAP_criterion_2} becomes equivalent to the \emph{maximum likelihood} (ML) criterion  \cite{Proakis} 
\begin{align}
  \hat{m}   &= \argmax_{1 \leq m \leq M} {\Pden(\rv \ | \ \sv_m)}.
  \label{eq:ML_criterion}
\end{align}

The likelihood pdf in \eqref{eq:ML_criterion} is obtained by considering the Jones vectors corresponding to $\rv$ and $\sv_m$,  noting that $\ket{r}-e^{\iota \theta}\ket{s_m} = \ket{n}$ by \eqref{eq:AWGNJoneschannelBis}, and averaging the pdf \eqref{gaussianDistro:eq} 
  over all $\theta$ \cite{Proakis}: 
\begin{align}\label{mainrsProb:eq}
  &\quad \Pden(\rv \ | \ \sv_m) \notag \\ 
  &= \frac{1}{(2 \pi \nvar)^\d} \int_0^{2 \pi} \exp\left({-\frac{\| \ket{r} - e^{\iota \theta}\ket{s_m} \|^2}{2 \nvar}}\right) \, \frac{d\theta}{2\pi} \notag \\
      & =   \frac{  1 }{(2 \pi \nvar)^\d}  \exp\left( -\frac{\braket{r}-\braket{s_m} }{2 \nvar} \right)  
              I_0\left( \frac{|\ip{r}{s_m}| }{\nvar}  \right),
\end{align}
 where we used the \emph{modified Bessel function of the first kind} of zero order
  \begin{equation}
    \label{eq:modBesselZero}
    I_0(x):= 
    \frac{1}{\pi} \int_0^{\pi} \exp\left({x \cos \theta}\right) \, d\theta.
  \end{equation}
\noindent Equality \eqref{mainrsProb:eq} is obtained by expanding in the exponent  (cf. \eqref{eq:AWGNStokesChannelIncoherent}): 
	\begin{align*}
		&\quad 
  {  \| \ket{r} - e^{\iota \theta}\ket{s_m} \|^2 } \notag \\
		& =  \braket{r}  + \braket{s_m}  -
		\underset{2 \cos(\theta-\theta_m) |\ip{r}{s_m}|}{\underbrace{2 \Re \left( e^{\iota (\theta-\theta_m) } |\ip{r}{s_m}| \right)}},    \addtag
		\label{eq:likelihood_exponent}
	\end{align*}	
where $\theta_m$ denotes the argument of $\ip{r}{s_m}$ (which is immaterial). The integration domain reduces to $[0,\pi]$ due to the symmetry of cosine.  

Based on  \eqref{mainrsProb:eq}, we can rewrite the {maximum likelihood} (ML) criterion \eqref{eq:ML_criterion} as \cite{McCloud}
\begin{align}
  \hat{m} 
  &= \argmax_{1 \leq m \leq M} |\ip{r}{s_m}|,
  \label{eq:optScheme}
\end{align}
where we used the monotonicity of $I_0(x)$ and the fact that $\ip{s_m}=1$. 

In particular,  in the Jones space, the 
 ML decision region ${\mathcal D}_m \subset \C^N$ for $\ket{s_m}$ is 
\begin{equation}
  \label{eq:Voronoi}
  {\mathcal D}_m = \left\{ \ket{r}: \   |\ip{r}{s_m}| \geq   |\ip{r}{s_{m'}}| \ \text{ for all $m' \neq m$}  \right\}.
\end{equation}

In passing, note that ${\mathcal D}_m$ viewed in the Stokes space is the  \emph{Voronoi cell} around $\sv_m$, and this is so irrespective of which applicable concept of distance,  $\dD$ or $\dStok$, is used (cf. Sec. \ref{sec:distances}).

The \emph{symbol error probability} is the expected probability of missing the right ML decision region, expressed by the following sum of integrals
with respect to the $2\d$-dimensional volume 
 in $\C^\d$: 
\begin{equation}
  \label{eq:errProb}
  \Pe = \sum_{m=1}^M P_m \sum_{m' \neq m} \int_{{\mathcal D}_{m'}}  \Pden(\rv \ | \ \sv_m) \, d \! \ket{r}. 
\end{equation}

In principle, $\Pe$ can be computed based on the channel model, but its analytic evaluation is impossible for all but the simplest constellations due to the complex geometry of the ML decision regions ${\mathcal D}_m$. Therefore, readily  computable analytic bounds on $\Pe$ are of value.


\subsection{Union Bound}

\noindent The general form of the union bound is  \cite{Proakis}
\begin{equation}
  \label{eq:UBgeneralForm}
  \Pe 
  \leq \sum_{m=1}^M P_m \sum_{m' \neq m}   \Peb^{m' | m} \ ,
\end{equation}
where $ \Peb^{m' | m}$ is the pairwise error probability of deciding on $\ket{s_{m'}}$ when $\ket{s_{m}}$ was sent in a binary fashion, while no other symbols are considered. Therefore,  
\begin{equation}
  \label{eq:mPrimeUnionBoundBis}
\Peb^{m' | m} =  \int_{{\mathcal D}^{ m'| m}} \Pden(\rv \ | \ \sv_m) \, d \! \ket{r}. 
\end{equation}
Above,  the \emph{pairwise error decision region}
${\mathcal D}^{ m'| m}$ is where 
$P_{m'} \Pden( \rv  \ | \ \sv_{m'} \ ) \geq P_{m} \Pden( \rv  \ | \ \sv_{m} \ )$, i.e., for equiprobable symbols,
\begin{align}
  \label{eq:mmprimeRegionBis}
                                            {\mathcal D}^{ m'| m} &= \left\{ \ket{r} \in \C^\d: \  |\ip{r}{s_{m'}}| \geq |\ip{r}{s_{m}}| \right\}.
\end{align}
Inequality \eqref{eq:UBgeneralForm} follows from the manifest inclusion ${\mathcal D}_{m'} \subset {\mathcal D}^{ m'| m}$. 

Our main result gives an explicit form for the terms in the union bound.

\begin{tcolorbox}
 \begin{thm}\label{main theorem Pbin}
   The binary error probability between two equiprobable non-orthogonal unit vectors $\ket{s_m}, \ket{s_{m'}} \in \C^\d$ is
\begin{empheq}[box=\fbox]{align}
     \label{mainExact:eq}
  \Peb^{ m'| m} 
  &= Q_1\left( \sqrt{\gamma_s}{\rho_-},\sqrt{\gamma_s}{\rho_+} \right) \notag \\
  &\quad -\frac{1}{2} \exp\left({-\frac{\gamma_s}{2}}\right) I_0\left(\frac{\gamma_s\gamma}{2}\right),
\end{empheq}
   where we defined the length parameters
\begin{subequations}
	\begin{align}
 \rho^2_\mp &:=\frac{1 \mp \delta}{2},\\ \delta &:= \sqrt{1 - \gamma^2},\\ \gamma & := |\ip{s_m}{s_{m'}}|>0,
\end{align}
\end{subequations}
    \nomenclature[$rho$]{$\rho^2_\mp$}{Length parameter, $\rho^2_\mp:= \frac{1\mp \delta}{2}$}
    \nomenclature[$delta$]{$\delta$}{Length parameter, $\delta:= \sqrt{1-\gamma^2}$}
 and $Q_1$ stands for the \emph{Marcum Q-function} of the first order defined by \cite{Proakis}
 \begin{equation}
   \label{eq:MarcumDef}
   Q_1(\aa, \bb) := \int_{\bb}^\infty x \exp\left({-\frac{x^2+\aa^2}{2}}\right) I_0(\aa x) \, dx.
 \end{equation}
\end{thm}
\end{tcolorbox}
The formulas extend to case of orthogonal signals, when $\gamma=0$  and one can fall back onto $Q_1(0, \bb)=\exp\left({-\frac{\bb^2}{2}}\right)$. To improve readability, we leave the proof of Theorem \ref{main theorem Pbin} to Appendix \ref{app:mainProofPes}.

For ease of reference, we instantiate \eqref{eq:UBgeneralForm} with \eqref{mainExact:eq} and record the following corollary. 
\begin{tcolorbox}
\begin{cor}
\label{cor:UBsumExact}
Given a  constellation of equiprobable unit vectors $(\ket{s_i})_{i=1}^M \in \C^N$ 
 the symbol error probability $\Pe$ is bounded as 
\begin{empheq}[box=\fbox]{align}
    \label{UBsumExact}
    \Pe &
    \leq \frac{1}{M} \sum_{m=1}^M \sum_{m' \neq m}  \Big[ Q_1\left( \sqrt{\gamma_s}{\rho_-},\sqrt{\gamma_s}{\rho_+} \right) \notag\\
     &\quad -\frac{1}{2} \exp\left(-\frac{\gamma_s}{2}\right) I_0\left(\frac{\gamma_s\gamma}{2}\right)\Big]. 
\end{empheq}
\end{cor}
\end{tcolorbox}

Below, we present asymptotic expressions that are valid for larger values of the symbol SNR $\gamma_s$ and are obtained by implementing the results reported in \cite{GilSeguraTemme2014}  (proven in \cite{Temme1993} and based on \cite{Temme1986, Goldstein1953}). The derivation of these formulas is given in Appendix  \ref{app:Asymptotics}.
\begin{tcolorbox}
\begin{cor}
  \label{asympt:cor}
  For large values of the symbol SNR $\gamma_s$, when $\gamma_s \rho_+\rho_- = \gamma_s \gamma/2$ is sufficiently large, we can use asymptotic expansions for $\Peb^{m' | m}$ and the zeroth and first order approximations are as follows: 
\begin{empheq}[box=\fbox]{align} \label{mainAsymptZero:eq}
    \Pbin 
    &\sim   \frac{1}{2} \sqrt{\frac{\gamma}{1-\delta}}        
         \erfc\left(\frac{\sqrt{\gamma_s}\sqrt{1-\gamma}}{\sqrt{2}} \right) \notag \\
    &\quad  - \frac{1}{2\sqrt{\pi \gamma \gamma_s}} \exp\left( {-\frac{\gamma_s(1-\gamma)}{2}}\right)
\end{empheq}
and
\begin{empheq}[box=\fbox]{align} 
  \label{mainAsympt:eq}
    &\quad \Pbin \notag \\
    &\sim   \left[ \frac{1}{2} \sqrt{\frac{\gamma}{1-\delta}}
               - \sqrt{\frac{1-\gamma}{\gamma}}  \frac{1}{8\sqrt{2}}\left(\frac{1+\delta}{\gamma} + 3 \right) \right]    \notag \\ 
    &\quad  \times \erfc\left(\frac{\sqrt{\gamma_s}\sqrt{1-\gamma}}{\sqrt{2}} \right) \notag \\
    &\quad + \frac{1}{8\sqrt{\pi}}\left[\sqrt{2} \sqrt{\frac{1-\gamma}{1-\delta}}\left(\gamma \gamma_s\right)^{-\frac{1}{2}}   - \left(\gamma \gamma_s\right)^{-\frac{3}{2}}   \right]  \notag \\
    &\quad  \times \exp\left({-\frac{\gamma_s(1-\gamma)}{2}} \right) . 
\end{empheq}
\end{cor}
\end{tcolorbox}

We note that the particular value of the asymptotic expressions in the corollary above is their ability to handle poorly separated vectors (with $\gamma \approx 1$). Pairs of vectors with small separation contribute the bulk of the $\Pe$.  Moreover, MVM constellations with  a given spectral efficiency per spatial degree of freedom (e.g., analogous to QPSK) have diminishing minimal distance with the increase of the number of cores/modes.
When $\gamma$ is not too close to $1$ and SNR is large, we have a simpler
asymptotic expression (with a straight forward derivation given in Appendix \ref{app:Asymptotics}): 
\begin{align}\label{mainAsymptOld:eq}
\boxed{     \Peb^{ m'| m}
  \sim \frac{1}{2} \frac{1}{\sqrt{\pi}}
  \sqrt{\frac{1+\gamma}{1-\gamma}}
  \frac{1}{\sqrt{\gamma}\sqrt{\gamma_s}} \exp \left[{-\frac{\gamma_s (1 - \gamma)}{2}}\right] .}
\end{align}

In any case,  the leading exponential asymptotics is
\begin{align}
\exp\left[-\frac{\gamma_s (1 - \gamma)}{2}\right] = \exp\left[-\frac{1}{2} \gamma_s \frac{\dD(\sv_m, \sv_{m'})^2}{2} \right],
\label{eq:asymptoticBehavior}
\end{align}
where $\dD(\sv_m, \sv_{m'})$ is the incoherent distance between $\sv_m$ and $\sv_{m'}$ (as defined in Sec. \ref{sec:distances}, ahead). This indicates that the distance $\dD$ is a natural way of expressing the proximity of the symbols in our context. 

We add that, if $\gamma=0$, \eqref{mainAsymptZero:eq}--\eqref{mainAsymptOld:eq} are not valid. However, then $\Peb^{m' | m}$ equals $\frac{1}{2}\exp(-\gamma_s /2)$ and is eclipsed by the terms with $\gamma >0$ in the sum giving the union bound \eqref{eq:UBgeneralForm}.  Even though \eqref{mainAsymptZero:eq} and  \eqref{mainAsympt:eq} work well even for moderately small values of $\gamma$, one can safely drop the terms with the smaller $\gamma \approx 0$.
 
Appendices \ref{app:mainProofPes} and \ref{app:Asymptotics} are devoted to the proofs of the Theorem \ref{main theorem Pbin} and the Corollary \ref{asympt:cor}, respectively. Sec. \ref{sec:results} shows comparisons of the union bound for  $\Pe$ obtained by using the above theoretical approximations and numerically computed Monte Carlo based values of $\Pe$ for several example constellations.

\subsection{Distance definitions}\label{sec:distances}

\noindent For the optimal geometric shaping of an MVM constellation, which is discussed in Sec. \ref{sec:GeometricShaping},  it is necessary to adopt some function of distance between constellation points.
The choice of distance function depends on the detection scheme and the nature of the dominant channel impairments. For the back-to-back performance evaluation of  optically-preamplified MVM direct-detection receivers, we consider that ASE noise is the dominant impairment. In this case, for equienergetic MVM constellations,  the suitable metric turns out to be the \emph{chordal Fubini-Study distance}, which is a special case of what we call below \emph{incoherent/direct-detection distance}.
We note that there are several arguments advocating naturality of this metric. Perhaps the strongest is based on the way it enters the previously derived asymptotic form of the union bound \eqref{eq:asymptoticBehavior} for the symbol error probability.  

One quick takeaway is that there is a better distance than the ordinary Euclidean Stokes distance, which is often the default choice and was also initially used in our computations.  Below, we define the \emph{incoherent/direct detection distance} and relate it to other common distance functions. 

\subsubsection{Coherent Distance }
\noindent In  Jones space $\C^\d$, we have the standard Euclidean distance between MVM symbols, which can be written as a function of the Hermitian inner product 
\begin{align}
	\label{eq:EuclidJOnesDist}
	\dC( \ket{s}, \ket{s'})&:=
	\|\ket{s}-\ket{s'}\| \notag\\
	&=\sqrt{ \ip{s}- 2\Re \ip{s}{s'}  + \ip{s'}}.
\end{align}
For unit vectors, \eqref{eq:EuclidJOnesDist} is  expressed in terms of the \emph{coherent angle}
$\thetaC \in [0, \pi]$ as 
\begin{equation}\label{eq:EuclidJOnesDistUnit}
	\dC( \ket{s}, \ket{s'})=\sqrt{2}\sqrt{1 - \cos \thetaC}, 
\end{equation}
where $\cos \thetaC :=\Re \ip{s}{s'}$.

We  refer to this distance as \emph{coherent distance}, since the probability of making a binary error between $\ket{s}$ and $\ket{s'}$ in a coherent receiver depends on $\dC$  in a natural way.

\subsubsection{Incoherent Distance}

\noindent In the case of incoherent detection, a transmitted MVM symbol is abstractly represented by a Jones vector up to phase, $e^{\iota \theta}\ket{s}$, with indeterminate $\theta \in [0, 2\pi)$. Mathematically, as long as $\ket{s}\neq0$, this is a circle in Jones space $\C^\d$.    
From this standpoint, one might guess that the natural distance between symbols $s$ and $s'$ is the minimum coherent Jones distance between the two circles:
\begin{align}
	\dD( \ket{s}, \ket{s'})&:= \min_{\theta,\theta'} \left\vert e^{\iota \theta}\ket{s} -  e^{\iota \theta'} \ket{s'} \right\vert \notag\\
	&= \sqrt{\|s\|^2 - 2 \left| \ip{s}{s'} \right|  + \|s'\|^2}.
\end{align}
When $\|s\|=\|s'\|=1$, we can use the \emph{incoherent/direct detection angle} $\thetaD \in [0,\pi/2]$: 
\begin{equation}\label{eq:ddDistUnit}
	\dD( \ket{s}, \ket{s'})=\sqrt{2}\sqrt{1 - \gamma},
\end{equation}
where $\gamma:= \cos \thetaD:= |\ip{s}{s'}|$. 

In this case, the distance  coincides with the {\it chordal Fubini-Study distance}
on the complex projective space.
In Appendices \ref{app:mainProofPes} and \ref{app:Asymptotics},  the incoherent cosine $\gamma = \cos \thetaD = |\ip{s}{s'}|$ of two normalized symbols under consideration  will make frequent appearance.

Of course, if only from $\Re \ip{s}{s'} \leq \vert \ip{s}{s'}\vert$, we have
\begin{equation}
	\dD(\ket{s}, \ket{s'}) \leq \dC(\ket{s}, \ket{s'}) \quad \text{ and } \quad  \thetaD \leq \thetaC.
\end{equation}
The loss of phase information degrades one's ability to distinguish symbols.

\subsubsection{Hilbert-Schmidt and Stokes Distance}


\noindent Another way to represent  incoherently-received symbols is with dyads ${\bf S}:=\dyad{s}$. 
Their natural ambient linear space is 
${\mathrm M}_{\d \times \d}(\C)$ of all $\d \times \d$ complex matrices, which can be used together with the Hilbert-Schmidt Hermitian inner product $\trace({\bf A}^\dagger {\bf B})$, where the operator $\trace(~)$ denotes the trace of a matrix.

The \emph{Hilbert-Schmidt distance} on ${\mathrm M}_{\d \times \d}(\C)$ is defined as  
\begin{align}
	\dHS({\bf A}, {\bf B})&:=\left\| {\bf A} - {\bf B}\right\|_{HS} \notag\\
	&= \sqrt{\trace({\bf A}^\dagger {\bf A}) - 2 \Re \trace({\bf A}^\dagger {\bf B}) + \trace({\bf B}^\dagger {\bf B})}.
\end{align}

Restricted to dyads, since $\left\| {\bf S} \right\|_{HS}^2 = \ip{s}^2$ and $\tr({\bf S}^\dagger{\bf S'})=|\ip{s}{s'}|^2$ is already real, we get 
\begin{equation}
	\dHS({\bf S},{\bf S'}) = \sqrt{\ip{s}^2 - 2|\ip{s}{s'}|^2 + \ip{s'}^2}.
\end{equation}
When $\|\ket{s} \|=\| \ket{s'} \|=1$, we could speak of \emph{Hilbert-Schmidt angle} and 
\begin{equation}
	\dHS({\bf S},{\bf S'})=\sqrt{2}\sqrt{1 - \gamma^2} \quad \text{where} \quad \gamma:= |\ip{s}{s'}|. 
\end{equation}

Traditionally, incoherently-received MVM symbols are represented by Stokes vectors $\sv \in \R^{\d^2-1}$, whose entries are the coefficients of the expansion of the \emph{trace neutralized dyad} ${\bf S}$, e.g., ${\bf S}-\frac{1}{\d}{\bf I}_N$ (assuming normalization $\ip{s}=1$), with respect to the Gell-Mann matrix basis \cite{Roudas_PJ_17}.
The Euclidean distance in  Stokes space, called \emph{Stokes distance},  coincides with the Hilbert-Schmidt distance up to scaling (having to do with the said trace adjustment and the conventions for the  Gell-Mann matrix basis): 
\begin{equation}
\label{stokDis:eq}
	\dStok(\sv,\sv') =  2C_{\d}\sqrt{1-\gamma^2}.
\end{equation}

The Stokes distance should be better suited for thermal noise-limited direct-detection receivers, not for their ASE noise-limited counterparts, which  have been the focus  of this work.

\section{Geometric Constellation Shaping}\label{sec:GeometricShaping}
A quintessential problem in digital communications systems is the optimal selection of signal sets to minimize the symbol error probability under various noise distributions and channel impairments. The term geometric constellation shaping means that the positions of constellation points
in the signal space are selected appropriately in order to minimize the  error probability. As a prototypical example, Foschini et al. \cite{Foschini1974OptimizationOT} numerically optimized  the shapes of two-dimensional signal constellations with arbitrary cardinality in the case of additive white Gaussian noise and coherent detection.  Extending this work to optical communications, Karlsson and Agrell \cite{karlsson2016multidimensional} investigated optimized power-efficient multidimensional modulation formats for coherent optical communications systems. For relatively small dimensions $N$, they used sphere-packing algorithms to optimize the constellation points. For larger dimensions, their design strategy was to select points from $N$-dimensional lattices \cite{karlsson2016multidimensional}.

In the case of SVM ($N=2$), geometric constellation shaping for equipower signal  sets (PolSK) was performed numerically initially by Betti et al. \cite{betti1990multilevel} by maximizing the minimum Euclidean distance among signals in Stokes space and then by Benedetto and Poggiolini \cite{benedetto1994multilevel} by using  the exact symbol error probability of $M$-ary PolSK modulation schemes for constellations of equipower signals as an objective function. To derive a formula for the symbol error probability, Benedetto and Poggiolini calculated the boundaries of the decision regions initially considering signal vectors in Stokes space that were placed at the vertices of a regular polyhedron inscribed within the Poincar\'e sphere, and then extended the analysis to generic equipower constellations with constellation points at the vertices of irregular polyhedra \cite{benedetto1994multilevel}. Optimum signal constellations for the case of $N=2$ and $M=$ 4, 8, 16 and 32 signals were derived \cite{benedetto1994multilevel}.  Kikuchi \cite{Kikuchi2014} used suboptimal 2D quaternary and cubic octary constellations for implementation simplicity. Morsy-Osman et al.\cite{MorsyOsman2019} designed intensity/polarization SVM constellations based on the face-centered cubic (FCC) lattice to achieve maximum packing density, assuming a thermal-noise-limited scenario and using the minimum Euclidean distance criterion.

 Our goal here is to  spread out  the MVM constellation points in the generalized Stokes space and, thus, improve the symbol error probability in a direct-detection-based link. 
 Since the adoption of symbol error probability as objective function leads to a computationally-intensive numerical optimization, a suitably-selected  simplified objective function is used instead.    The gradient-descent method \cite{boyd2004convex} is used for the minimization of the simplified objective function.
 
 To facilitate calculations, we consider an objective function from  electrostatics \cite{wiki:Thomson}, wherein the constellation points are assumed to be identical charges on the surface of a perfectly conducting Poincar\'e hypersphere. Starting from given initial positions, the charges are allowed to equilibriate under the action of Coulomb forces. In other words, we recast the original three-dimensional Thomson problem \cite{wiki:Thomson} to higher-dimensional Stokes space.	This adaptation requires constraining the $M$ constellation points to a $(2N-2)$--dimensional manifold  due to the relationships \eqref{eq:jv}, \eqref{eq:StokesVectorsdf} relating the higher-dimensional Jones and Stokes spaces  \cite{Antonelli:12, Roudas_PJ_17, Roudas_JLT_18}. 
 
 It is worth saying a few words here about the extensive literature on the Thomson problem. Since the original publication of the problem by  J. J. Thomson’s in 1904,  numerous papers were written on this topic and its variants. Saff and Kuijlaars \cite{Saff} give a comprehensive survey of the literature in the two dimensional case $N=2$, with an emphasis on the case when $M$ is large. Global minima for the Thomson Problem for $N=2$ are posted on the Cambridge website \cite{Cambridge_Thomson_website}. 

The function \lstset{language=Mathematica}{SpherePoints[n]}
in Wolfram \textit{Mathematica}\textregistered \, \cite{kogan2017} gives the positions of $n$ approximately uniformly distributed points on the surface of the $S^2$ unit sphere in three dimensions, with exact values for certain small $n$ and a spiral-based approximation for large $n$ \cite{kogan2017}.

Closely related to Thomson's problem is the Tammes problem whose goal is to find the arrangement of $M$ points on a unit sphere which
maximizes the minimum distance between any two points. Jasper et al. \cite{jasper2019game} studied the Tammes problem in the complex projective space and maintain a website listing the current best-known numerical approximations \cite{Math_Colorado_State_website}.

%
%
%
%
%
%

\subsection{Gradient computation}

\noindent Consider a perfectly conducting Poincar\'e hypersphere with identical charges at the positions of the constellation points. As charges repel each other with Coulomb forces, they move on the surface of the Poincar\'e hypersphere until they reach an equilibrium distribution with minimum potential energy. 

The electrostatic potential energy $\Omega(d_{ij})$ of two charges $i,j$ separated by a distance $d_{ij}$ is inversely proportional to their distance $\Omega(d_{ij})\sim d_{ij}^{-1}$. The total electrostatic potential energy $U$ of a system of $M$ charges can be obtained by calculating the  potential energy $\Omega(d_{ij})$ for each individual pair of charges $i,j$ and adding the  potential energies for all distinct combinations of charge pairs
\begin{equation}
  \label{eq:ColumbPot:eq}
  U = \sum_{i=1}^{M}\sum_{j=i+1}^{M}\Omega(d_{ij}).
\end{equation}

The distances $d_{ij}$ can be calculated in terms of the corresponding unit Jones vectors $\ket{s_i} \in \C^\d$, $i=1, \ldots, M$. 
\begin{equation}
\label{dijDistGenFunction:eq}
d_{ij} = \psi\left(\lvert \ip*{s_i}{s_j} \rvert^2\right) = \psi \left( \gamma^2 \right)
\end{equation}
 We leave the function $\psi$ unspecified for now to allow use of various distances between Stokes vectors (cf. Sec. \ref{sec:distances}).
 
To compute the gradient, we first assume that the Jones vectors depend on a certain parameter $t$ and compute  
\begin{align}
  \label{partDerRec:eq}
     \frac{\partial U}{\partial t} 
    &=   \sum_{i<j} \Omega'(d_{ij})  \frac{\partial d_{ij}}{\partial t} \notag\\
    &=   \sum_{i<j}  \Omega'(d_{ij})  \psi'\left( |\ip*{s_i}{s_j}|^2 \right) \frac{\partial |\ip*{s_i}{s_j}|^2}{\partial t} \notag\\
  &=  \sum_{i<j} \Omega'(d_{ij}) \psi'\left( |\ip*{s_i}{s_j}|^2 \right) \cdot 2 \Re\left(  \ip*{s_j}{s_i} \ip{\frac{\partial s_i}{\partial t}}{s_j} \right.\notag\\
    &\qquad \left. +   \ip*{s_i}{s_j} \ip{\frac{\partial s_j}{\partial t}}{s_i} \right),
\end{align}
where we used multilinearity to evaluate 
\begin{align}
  \frac{\partial |\ip*{s_i}{s_j}|^2}{\partial t}  &=  \frac{\partial}{\partial t} \ip*{s_j}{s_i} \ip*{s_i}{s_j} \notag\\
    &= \ip{\frac{\partial s_j}{\partial t}}{s_i} \ip*{s_i}{s_j} +  \ip*{s_i}{s_j} \ip{\frac{\partial s_j}{\partial t}}{s_i} .
\end{align}

Taking $t$ to be the real and imaginary parts of $s_{im}=x_{im} + \iota y_{i,}$, the components of the gradient of $U$ are found as   
\begin{align}
  \frac{\partial U}{\partial x_{im}}
  &=  \sum_{j: \ j \neq i} \Omega'(d_{ij}) \psi'\left( |\ip*{s_i}{s_j}|^2 \right)
    2 \Re\left(  \ip*{s_j}{s_i} s_{jm} \right)
\end{align}
and
\begin{align}
  \frac{\partial U}{\partial y_{im}}
  &=  \sum_{j: \ j \neq i} \Omega'(d_{ij}) \psi'\left( |\ip*{s_i}{s_j}|^2 \right)
    2 \Im \left(  \ip*{s_j}{s_i} s_{jm} \right).
\end{align}
To state the end result, the gradient\footnote{N.B.: This is not a complex derivative as $U$ is not necessarily analytic.} $\nabla U$ is the vector of real and imaginary parts of the (complex) vector
$\left( \frac{\partial U}{\partial s_{im}} \right)_{i,m} \in \C^{M\times\d}$ given by 
\begin{align}
  \label{mainGradientFormula:eq}
  \frac{\partial U}{\partial s_{im}}
  &=   2 \sum_{j:\ j \neq i} \Omega'(d_{ij}) \psi'\left( |\ip*{s_i}{s_j}|^2 \right)
    \ip*{s_j}{s_i} s_{jm}.
\end{align}

\subsection{Example: Coulomb Potential}

\noindent In the following, we adapt the three-dimensional Thomson problem \cite{wiki:Thomson} to the generalized Stokes space. It is true that the use of the electrostatic potential energy as an objective function in lieu of the symbol error probability  is not justified by the underlying physics of the problem under study. Nevertheless, as shown in Fig. \ref{fig:potential comparison}, the minimization of the electrostatic potential yields nearly optimal results that are very close to the ones obtained by minimizing the symbol error probability.    

For the Thomson problem, we use  the Euclidean distance $\dStok$ in the Stokes space,  per \eqref{stokDis:eq},  
 so that  \eqref{dijDistGenFunction:eq} is written as
\begin{equation}
  \psi(t) :=  2C_{\d}\sqrt{1-t}, 
 \end{equation}
 where now $t=\gamma^2$.
 
 From $\psi(t)^2 = -4C_\d^2t + \text{Const}$, we get 
  $\psi'(t) = -2C_\d^2 \psi(t)^{-1}$, so 
\begin{align}
  \frac{\partial U}{\partial s_{im}}
  &=   -4C_\d^2 \sum_{j:\ j\neq i} \Omega'(d_{ij}) \psi\left( |\ip*{s_i}{s_j}|^2 \right)^{-1} \ip*{s_j}{s_i} s_{jm} \notag \\
  &=   -4C_\d^2 \sum_{j:\ j\neq i} \Omega'(d_{ij}) d_{ij}^{-1} \ip*{s_j}{s_i} s_{jm}.
\end{align}
Furthermore, for the case of electrostatic Coulomb forces acting in the Stokes space, we have the inverse distance potential 
\begin{equation}
  \Omega(d_{ij}) = d_{ij}^{-1}, \quad \Omega'(d_{ij}) = -d_{ij}^{-2}.
\end{equation}
Thus, instantiating \eqref{mainGradientFormula:eq} yields
\begin{equation}
\boxed{\frac{\partial U}{\partial s_{im}}
  =   4C_\d^2 \sum_{j:\ j\neq i} d_{ij}^{-3}  \ip*{s_j}{s_i} s_{jm}.}
  \label{eq:gradient_Thomson}
\end{equation}

 \subsection{Numerical details}
 \label{subsection: numerical details}


 
 
 \noindent We developed an efficient, partially-compiled Mathematica code implementing the gradient-descent optimization algorithm for arbitrary potentials. This implementation  is adequately fast on a personal computer to enable the design of MVM constellations with up to $M=1024$ points for up to $N=8$  spatial degrees of freedom (SDOFs) (see Fig. \ref{fig:SEplot_MN_MVM}).

 The gradient-descent optimization algorithm starts with either a randomly-generated constellation or a small random perturbation of a deterministic constellation.
 To give an example, consider the following deterministic constellation of Jones vectors: Their first component is set equal to unity, while their remaining $N-1$ components take all possible combinations of values in $\{\pm 1, \pm \iota\}$. Finally, the Jones vector length is normalized to unity. This process yields an MVM constellation with $M=4^{\d-1}$ vectors. We call it the \emph{standard constellation}. 
 Mathematically, it represents 
  the orbits of the vertices of a  hypercube
in $\C^\d$ under the circle action by the 
 phase rotation. {For this reason, it is possible to refer to it as the \emph{standard reduced hypercube constellation} or the \emph{standard reduced Jones hypercube constellation}.}

\section{Bit-to-symbol mapping}\label{sec:b2s}
Once we have geometrically optimized a constellation $(\ket{s_i})_{i=1}^M$ to reduce 
 the symbol errors, we seek to minimize bit errors by optimizing the bit-to-symbol mapping. In commonly-used modulation formats, such as $M$-PAM, $M$-PSK,  $M$-QAM on a square lattice, and its generalization to cubic lattices of any dimension, this task is achieved via \emph{Gray coding} \cite{AgrellGrayCoding}. Unfortunately, in  general, no such labeling readily exists for the MVM format.

Given a bit encoding $({\bf b}_m)_{m=1}^M$, where the length of each bit sequence ${\bf b}_m$ is $k=\log_2(M)$, we use the union bound \eqref{eq:UBgeneralForm} 
 to find that the \emph{average bit error probability} ${P}_{e|b}$  at a symbol SNR $\gamma_s$  is bounded by
\begin{equation}
    \label{eq:UBbitErrorProb}
    {P}_{e|b} \leq \frac{1}{kM} \sum_{m=1}^M \sum_{m'\neq m} \Pbin h_{mm'},  
\end{equation}
where $h_{mm'}$ denotes the \emph{Hamming distance}\nomenclature[$hmm$]{$h_{mm'}$}{Hamming distance between bit sequences ${\bf b}_m$ and ${\bf b}_{m'}$} between ${\bf b}_m$ and ${\bf b}_{m'}$. This is based on the observation that the expected number of bit errors corresponding to mistakenly receiving $\ket{s_{m'} }$ when $\ket{s_m}$ was transmitted is $\Pbin h_{mm'}$. Using the right side of  \eqref{eq:UBbitErrorProb}, we arrive at an objective function $\xi$ for evaluating various bit encodings:
\begin{align}
    \label{eq:SimulatedAnnealObjective}
    \xi &= \xi\left( (\ket{s_m} )_{m=1}^M, ({\bf b}_m)_{m=1}^M, \nvar \right) \notag \\
    &= \frac{1}{kM}  \sum_{m=1}^M \sum_{m'\neq m} \Pbin  h_{mm'}.
\end{align}
\nomenclature[$xi$]{$\xi$}{Simulated annealing objective function}

Finding a bit encoding $({\bf b}_m)_{m=1}^M$ that minimizes $\xi$ at a symbol SNR $\gamma_s$ serves as a proxy for minimizing bit errors for a given constellation and, thus, finding the optimal bit-to-symbol mapping. With $M!$ possible encodings, the sheer number of combinations prohibits brute-force solutions for all but the smallest constellations. This optimization problem can be viewed as a type of \emph{Quadratic Assignment Problem} \cite{Garey1979}, i.e., the optimal assignment of $\{1, 2, \dots, M\}$ (in binary) to $(\ket{s_m})_{m=1}^M$ with pairwise distances given by $\frac{1}{kM}\Pbin$ and pairwise weights given by Hamming distances $h_{mm'}$. Quadratic Assignment Problems are known to be NP-hard \cite{SahniPcomplete} and encompass the classical \emph{Traveling Salesman Problem} as a special case. 

Given these rapidly scaling combinatorics, we turn to numerical minimization. In particular, \emph{simulated annealing} has a long history of use for combinatorial optimization problems \cite{KirkpatrickSimulAnneal}, and lie within the broader class of \emph{Metropolis-Hastings algorithms}. Inspired by metallurgy, simulated annealing algorithms work by stochastically exploring the search space, helping prevent the algorithm from becoming entrapped near local minima. 

Our implementation begins with an initial bit-to-symbol mapping $({\bf b}_m)_{m=1}^M$ (either randomly selected or the current best known encoding) and a sequence of \emph{temperatures} $(T_n)$ per a selected \emph{cooling schedule} \cite{KirkpatrickSimulAnneal}.
In each iteration, a new candidate encoding $({\bf b}_m')_{m=1}^M$ is generated by randomly swapping the bit encodings for two symbols. We then compare $\xi \left( (b_m')_{m=1}^M \right)$ against $\xi\left( ({\bf b}_m)_{m=1}^M \right)$. If $\xi \left( ({\bf b}_m')_{m=1}^M \right) < \xi\left( ({\bf b}_m)_{m=1}^M \right)$, then $({\bf b}_m')_{m=1}^M$ is automatically accepted. Otherwise, $({\bf b}_m')_{m=1}^M$ is probabilistically accepted or rejected  by comparing $\exp\left( \left( \xi\left(({\bf b}_m)_{m=1}^M \right) - \xi\left( ({\bf b}_m')_{m=1}^M \right) \right)/T_n \right)$ against a uniformly randomly generated value in $[0,1]$. The initial high temperatures values give a higher probability of accepting a candidate encoding $({\bf b}_m')_{m=1}^M$ in order to explore the search space, while final low temperatures exploit local optimizations.

Implementing a simulated annealing optimization algorithm inherently requires significant tuning of parameters. Choices such as initial and final temperatures, cooling schedule, and number of iterations must all be carefully selected for the specific problem in order to properly balance exploration versus exploitation. After an investigation of  various cooling schedules, we established that a classic exponential cooling schedule of $T_n = \alpha^n T_0$ was well-suited to this problem. 
With further experimentation, we found that setting the initial temperature $T_0$ as the standard deviation of $\xi$ for a random sample of bit encodings gave acceptable performance across a wide range of constellation sizes $M$, without the need for extensively tuning this parameter. 

We remark that an efficient implementation will leverage that the constellation $(\ket{s_m})_{m=1}^M$ is static and hence the $\Pbin$ terms in \eqref{eq:SimulatedAnnealObjective} need only be computed once at the outset and then stored for all future evaluations of $\xi$.

Finally, we note that $h_{mm'}$ is trivially bounded by $k$ for all $m \neq m'$. Hence the performance increase that can possibly be achieved by optimizing the bit-to-symbol mapping $({\bf b}_m)_{m=1}^M$ is limited by a factor of $k$ (cf. Fig. \ref{fig:BERvRandom}), in contrast to the several orders of magnitude of performance that can be obtained by geometrically optimizing the constellation $(\ket{s_m})_{m=1}^M$ (cf. Fig. \ref{fig:potential comparison}). Therefore, the allocation of computation time when generating an $(N,M)$-MVM format should place greater emphasis on geometric optimization, while not completely neglecting to optimize the bit-to-symbol mapping.

\section{Results and discussion}\label{sec:results}
In Sec. \ref{sec:Pes}--\ref {sec:GeometricShaping}, we derived an upper bound for the symbol error probability of $(N,M)$-MVM and discussed accelerated geometric constellation shaping in the generalized Stokes space using an electrostatic analog (i.e., an extension of the Thomson problem to higher dimensions). In Sec. \ref {sec:b2s}, we proposed a method to optimize the bit-to-symbol mapping of arbitrary MVM constellations using simulated annealing. In this section, we navigate the reader through the steps of the  formalism presented in Sec. \ref{sec:Pes}--\ref{sec:b2s}  by providing illustrative examples for specific $N$, $M$. 

\subsection{Constellation design}
\noindent As a starting point, to develop some physical intuition by visualization, we consider  constellation shaping and  bit-to-symbol mapping in the three-dimensional Stokes space.

\begin{figure}[!htb]
	\centering
	\includegraphics[width=2in]{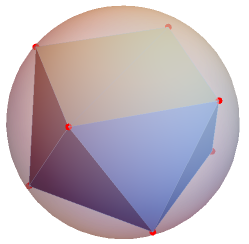}
	\caption{Square antiprism (Conditions: $N=2$, $M=8$).}
	\label{fig:polyhedronN_2_M_8.png}
\end{figure}

We  first examine the optimal distribution of eight points on the surface of the Poincar\'e sphere $S^2$. From \cite{benedetto1994multilevel}, we know that the optimal constellation corresponds to a square antiprism inscribed in the sphere as shown in Fig. \ref{fig:polyhedronN_2_M_8.png} (rather than a cube as proposed by \cite{Kikuchi2014}). We want to test whether the solution of the Thomson problem using the method of gradient descent coincides with the solution of \cite{benedetto1994multilevel}.

\begin{figure}[!htb]
	\centering
	\includegraphics[width=3in]{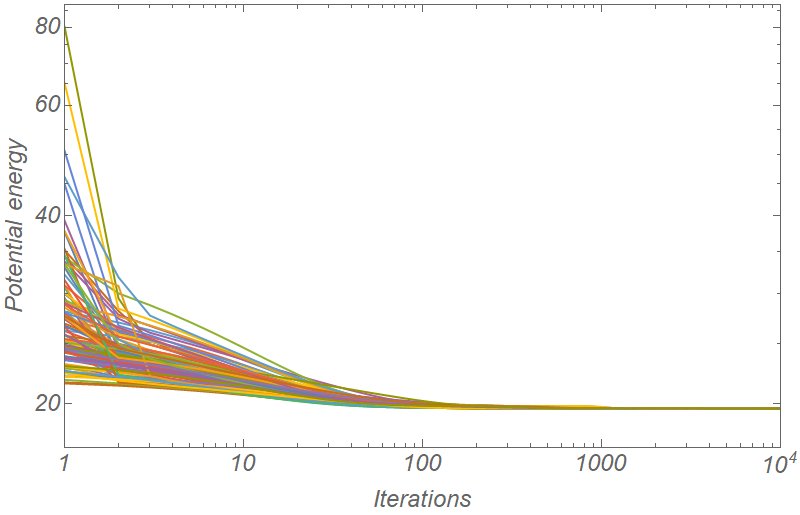}
	\caption{Thomson algorithm convergence for 100 distinct initial configurations (Conditions: $N=2$, $M=8$). After roughly 1,000 iterations, all cases have converged to square antiprisms.}
	\label{fig:ThomsonConvergence_N2_M8_N2.png}
\end{figure}

Fig. \ref{fig:ThomsonConvergence_N2_M8_N2.png} shows the evolution of the potential energy given by \eqref{eq:ColumbPot:eq} as a function of the number of gradient descent iterations associated with 100 different random initial configurations of $M=8$ point charges on $S^2$. After about 1,000 iterations, all cases converge to  essentially identical square antiprisms (up to arbitrary 3D rotations), like the one shown in Fig. \ref{fig:polyhedronN_2_M_8.png}. 

Close inspection reveals that the Euclidean distances between constellation points provided by the solution of the Thomson problem using the method of gradient descent in Fig. \ref{fig:ThomsonConvergence_N2_M8_N2.png} are slightly different from the ones provided by \cite{benedetto1994multilevel}. Actually, the constellation of \cite{benedetto1994multilevel} is unstable from an electrostatic point-of-view. In other words, if the constellation of \cite{benedetto1994multilevel} is provided as an initial configuration for the Thomson problem, the gradient in \eqref{eq:gradient_Thomson} of the potential energy in \eqref{eq:ColumbPot:eq} is non-zero, and, therefore, the constellation points experience Coulomb forces that move them to slightly different final positions. The same holds if one uses as initial guesses for the Thomson problem various point configurations provided by the minimization of alternative cost functions, e.g., for the Tammes problem \cite{jasper2019game}.

In conclusion, the polytopes provided by the minimization of different cost functions  for $N=2, M=8$ correspond to slightly different square antiprisms. For practical engineering purposes, however, we consider that these differences among various constellation configurations are immaterial and that the numerical solution of the Thomson problem using the method of gradient descent provides sufficient optimization effectiveness at low computational cost.
 
\begin{figure}[!htb]
	\centering	\includegraphics[width=2in]{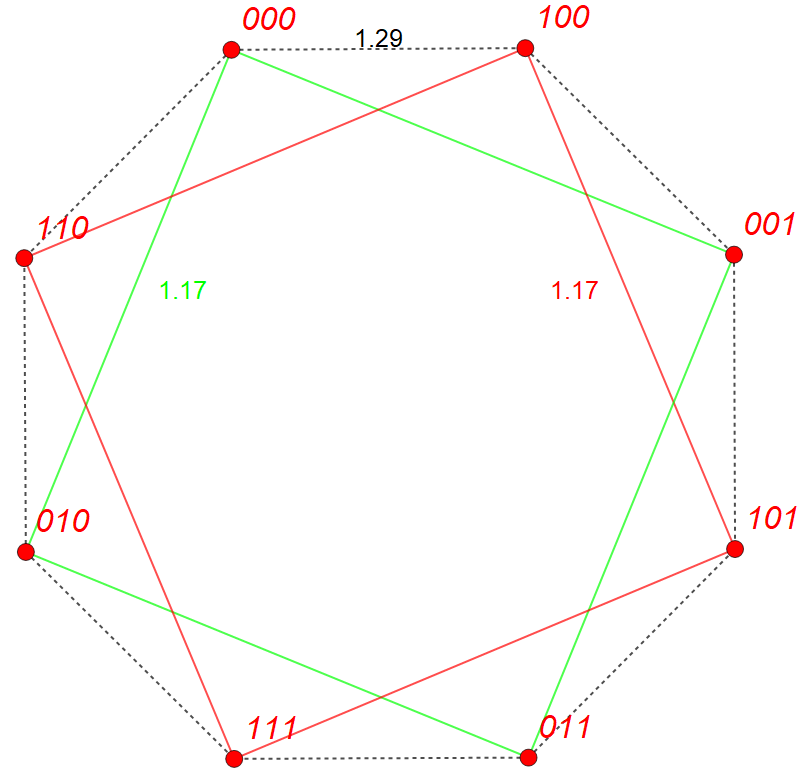}
	\caption{Bit-to-symbol mapping for the square antiprism (Conditions: $N=2$, $M=8$). The red, green, and black edges have lengths of approximately $1.17$, $1.17$, and $1.29$, respectively.}
	\label{fig:Graph_ThomsonqubicAntiprism_N2_M8.png}
\end{figure}

Next, we  shift our focus to the optimal bit-to-symbol mapping for the square antiprism. To facilitate visualization, we can represent the configuration of constellation points on the surface of the Poincar\'e sphere by a two-dimensional graph whose vertices represent the constellation points and its edges represent closest neighbors. For the case of the square antiprism of Fig. \ref{fig:polyhedronN_2_M_8.png}, we obtain the graph shown in Fig. \ref{fig:Graph_ThomsonqubicAntiprism_N2_M8.png}. The two square faces on opposite sides of the square antiprism are shown in red and green respectively, and the edges interconnecting them are shown with dotted black lines. The two square faces have sides equal to 1.17 and the edges interconnecting them are 1.29 long.

In Gray coding, closest neighbors at distance $1.17$ are assigned binary words that differ in only one bit, i.e., they have  a Hamming distance of one. Since each vertex in Fig. \ref{fig:Graph_ThomsonqubicAntiprism_N2_M8.png} has only two closest neighbors belonging to the same square face, it is straightforward to Gray label the vertices of the square faces using all binary words of three bits. For instance, one can Gray code the green square using the binary words with their most significant bit (msb) equal to zero and then use the remaining binary words with their most significant bit equal to one for the red square. The  proposed bit-to-symbol mapping in Fig. \ref{fig:Graph_ThomsonqubicAntiprism_N2_M8.png} is just one of  many possible Gray mappings.

However,  since the second-closest neighbors at distance $1.29$ are not very different distance-wise compared to the first neighbors at distance $1.17$, we have to take into account that erroneous symbol decisions can lead to second-closest neighbors with significant probability. 
 The proposed bit-to-symbol mapping in Fig. \ref{fig:Graph_ThomsonqubicAntiprism_N2_M8.png} offers almost all the benefits of Gray coding.
Each symbol error leads to 3 neighboring nodes that differ by one bit and to only one neighboring node that differs by two bits.

In  this particular case, the problem of assigning binary words to constellation points in order to minimize the bit error probability can be solved manually as follows: starting with the green square, we go into the clockwise direction and  
assign bits to symbols using all Gray words of zero msb. Then, starting from the vertex between 000 and 001, we trace the red square into the counterclockwise direction and  
assign bits to symbols using all Gray words of unit msb. We verified that the solution obtained via the simulated annealing algorithm is indeed the one found manually in Fig. \ref{fig:Graph_ThomsonqubicAntiprism_N2_M8.png}. This is evidence that the simulated annealing algorithm  performs adequately. 

In order to further validate bit-to-symbol mappings provided by our simulated annealing algorithm, we ran benchmarking tests on constellations that admit Gray coding \cite{grayCode} (e.g., M-PSK and M-QAM). Our implementation of the simulated annealing algorithm displayed strong performance in these tests, often finding the global minimum for small constellation sizes. 

For larger $M$, the computational complexity of assigning binary words to constellation points in order to minimize the bit error probability grows exponentially. Let us see why that is: 
There are $M!$ ways that we can assign $M$ words of $k$ bits to the $M$ nodes. Using the dominant term in Stirling's approximation for factorials, we see that $M!\sim M^{M}e^{-M}$ for $M \gg 1$. Computing the objective function for all possible arrangements and selecting the bit-to-symbol mapping that yields the global minimum is clearly computationally prohibitive for large values of $M$. Simulated annealing can be used to solve such combinatorial optimization problems. 
While it may not find globally optimal solutions, evidence from tests performed on small constellation sizes suggests that simulated annealing can produce bit-to-symbol mappings that are sufficiently nearly-optimal.
 

We continue by examining constellation shaping and  bit-to-symbol mapping in higher-dimensional Stokes spaces based on the physical intuition provided by  
the three-dimensional Stokes space.

\begin{figure}[ht] 
\captionsetup[subfigure]{justification=centering}
	\begin{subfigure}[b]{0.25\linewidth}
		\centering
		\includegraphics[width=0.75\linewidth]{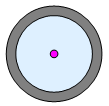}  
		\caption{} 
		\label{fig7:a} 
	\end{subfigure}
	\begin{subfigure}[b]{0.25\linewidth}
		\centering
		\includegraphics[width=0.75\linewidth]{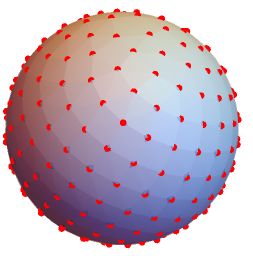}  
		\caption{} 
		\label{fig7:b} 
	\end{subfigure} 
	\begin{subfigure}[b]{0.25\linewidth}
		\centering
		\includegraphics[width=0.75\linewidth]{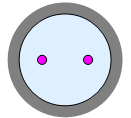} 
		\caption{} 
		\label{fig7:c} 
	\end{subfigure}
	\begin{subfigure}[b]{0.25\linewidth}
		\centering
		\includegraphics[width=0.65\linewidth]{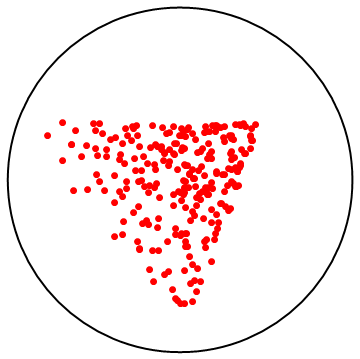}  
		\caption{} 
		\label{fig7:d} 
	\end{subfigure} 
	\caption{Illustration of optimized constellations (b), (d),  with $M$=256 points for (a) SMF ($N$=2); and (c) dual-core MCF ($N$=4), respectively.}
	\label{fig:constellations}
\end{figure}

Examples of optimized  constellations for $N=2$ and $N=4$ and $M=256$ are shown in Fig. \ref{fig:constellations}(b), (d), respectively.

\begin{figure}[!htb]\centering
	\includegraphics[width=2.5in]{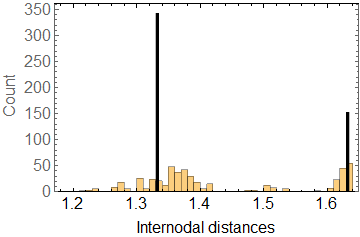}
	\caption{Histograms of internodal distances. (Yellow: Thomson problem; Black: Tammes problem \protect\cite{jasper2019game}).  (Conditions: $N=4$, $M=32$).} 
	\label{fig:HistogramComparison_M32_N4.png}
\end{figure}

\begin{figure}[!htb]\centering
	\includegraphics[width=2.5in]{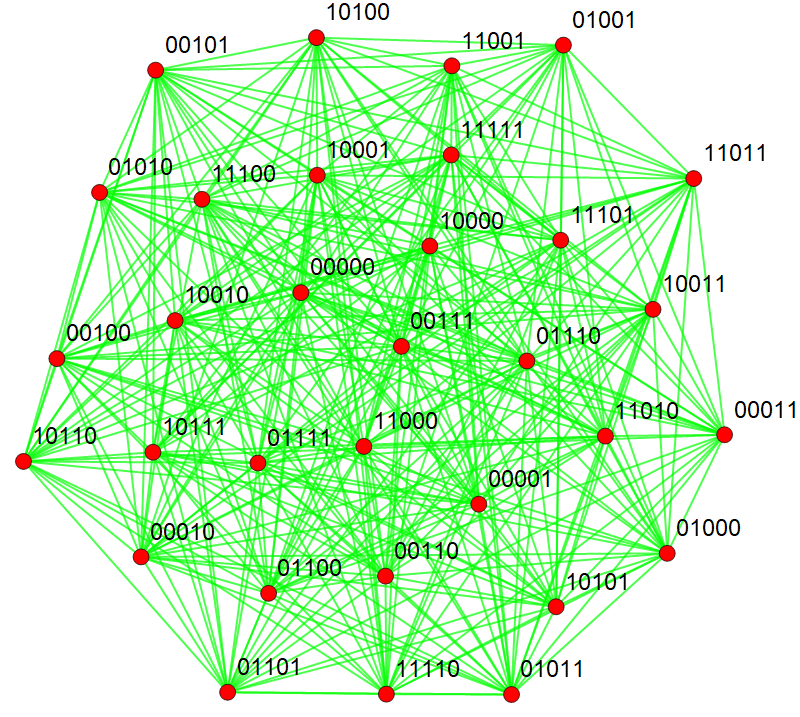}
	\caption{Constellation graph and bit-to-symbol mapping  (Conditions: $N=4$, $M=32$).} 
	\label{fig:GraphSloanes4x32_2.png}
\end{figure}

\begin{figure}[!htb]\centering
	\includegraphics[width=2.5in]{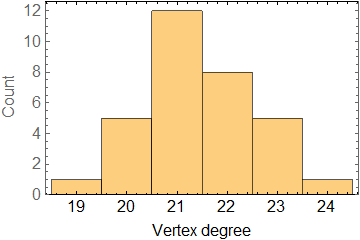}
	\caption{Histogram of vertex degrees for the constellation graph of Fig. \ref{fig:GraphSloanes4x32_2.png} (Conditions: $N=4$, $M=32$).} 
	\label{fig:HistogramVertexDegrees4x32.png}
\end{figure}

To illustrate the difficulties of bit-to-symbol mapping in higher-dimensional Stokes spaces, let us take a closer look at the optimized MVM constellation for $N=4$, $M=32$. The histograms of internodal distances for the Thomson problem and the Tammes problem \cite{Math_Colorado_State_website} are shown in Fig. \ref{fig:HistogramComparison_M32_N4.png}. Notice that the constellations found by solving these two problems are not identical. For instance, there are 343 closest neighbor pairs at distance 1.33 in the Tammes problem, whereas the Thomson problem gives a continuum-like distribution of internodal distances in the range 1.1-1.6 for the closest neighbors. Choosing the Tammes problem solution due to its high degree of symmetry, we make a 2D graph of the 32 vertices with edges interconnecting closest neighbors only (Fig. \ref{fig:GraphSloanes4x32_2.png}). Since the average vertex degree in the graph is 21 (Fig. \ref{fig:HistogramVertexDegrees4x32.png}), it is obvious that Gray coding cannot be applied. For 32 constellation points, the number of possible codings is $32!\approx 2.6\times 10^{35}$, so a brute force optimization by exhaustive enumeration is impossible. 
The bit-to-symbol mapping given by simulated annealing is shown in Fig.  \ref{fig:GraphSloanes4x32_2.png}.

\subsection{Validity of the error probability upper bounds}
\begin{figure}[!htb]
	\centering
	\includegraphics[width=3in]{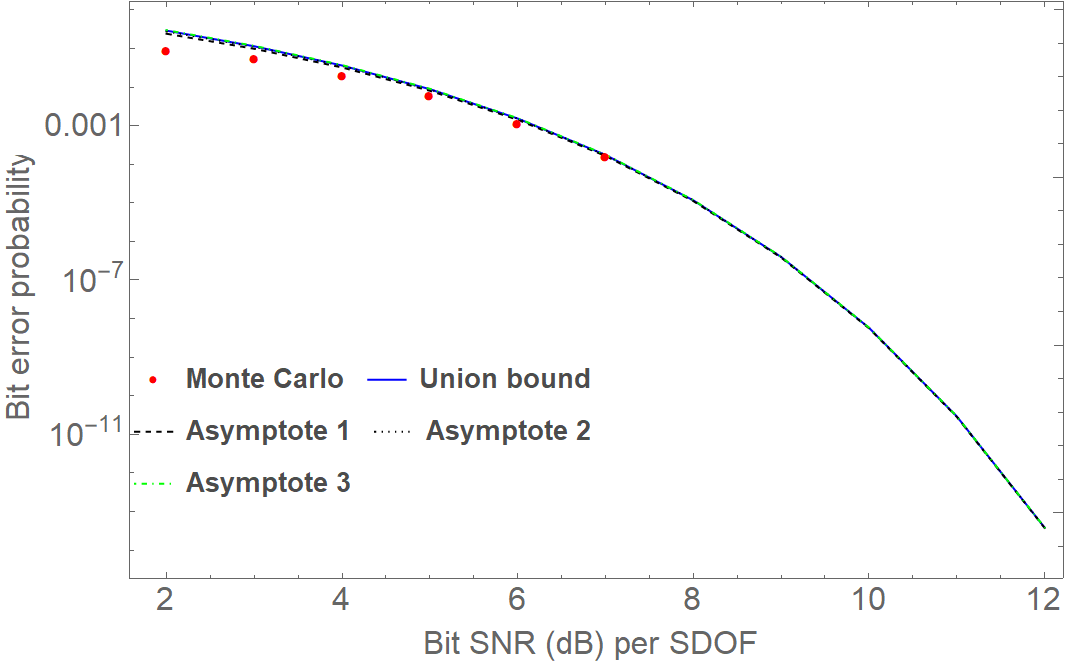}
	\caption{Bit error probability vs bit SNR (dB) per SDOF (Points: Monte Carlo simulation; Blue line: union bound (Corollary \ref{cor:UBsumExact}); dashed lines: asymptotics (Corollary \ref{asympt:cor})(Conditions: $N=8$, $M=64$).}
	\label{fig:Aymptotics_peb_vs_bit_snr_N_8_M_64_SpherePoints.png}
\end{figure}

The symbol error probability for equienergetic signals is bounded by using  the analytical union bound of Corollary \ref{cor:UBsumExact}. We want to gain insight into the validity and the tightness of this bound 
 at various bit SNRs. In Fig. \ref{fig:Aymptotics_peb_vs_bit_snr_N_8_M_64_SpherePoints.png}, we check the validity of Corollary \ref{cor:UBsumExact} and the asymptotic expressions of Corollary \ref{asympt:cor} 
 by Monte Carlo simulation. 
  We observe that the union bound is asymptotically tight and spot-on for bit error probabilities below the order of $10^{-3}$. Otherwise,   the union bound  overestimates larger error probabilities due to the significant overlap between the pairwise decision regions. 

\subsection{Impact of bit-to-symbol mapping}
\begin{figure}[!htb]
	\centering
	\includegraphics[width=3in]{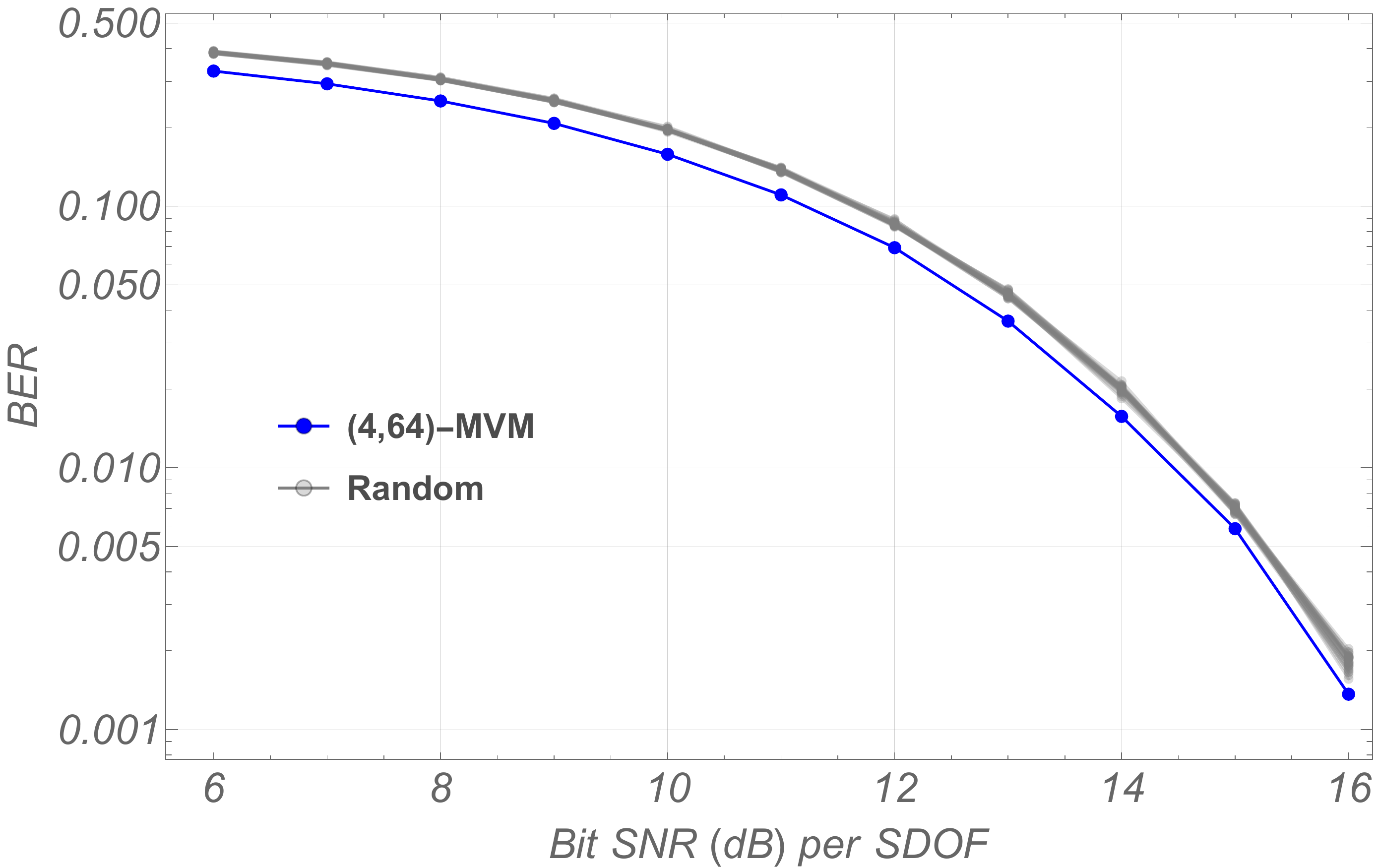}
	\caption{Bit error rate (BER) versus bit SNR per SDOF \nomenclature[$sdof$]{SDOF}{Spatial degrees of freedom}  for the optimized bit-to-symbol mapping (in blue) and randomized bit-to-symbol mappings (in gray).}
	\label{fig:BERvRandom}
\end{figure}

Fig.  \ref{fig:BERvRandom} compares the bit error rates (computed using Monte Carlo simulation) of the optimized bit-to-symbol mapping provided by simulated annealing (in blue) against multiple randomized encodings (in gray) for the same $(4,64)$-MVM constellation. We observe a performance gain of the optimized encoding over randomized encodings across a wide range of bit SNRs. In particular, we note that our bit-to-symbol mapping optimization requires a concrete choice of noise level $\sigma^2$ in defining the objective function of  \eqref{eq:SimulatedAnnealObjective} for simulated annealing. Hence it is possible that the suitability of an encoding might change with the noise level, requiring different optimizations for different noise levels. However, Fig. \ref{fig:BERvRandom} shows that a bit-to-symbol mapping optimized at one noise level (in this case, a bit SNR per SDOF of $10$ dB) performs well across a range of SNRs, showing that this concern is immaterial in practice. 


\subsection{Potential selection for constellation optimization}

\begin{figure}[!htb]
    \centering
    \includegraphics[width=3in]{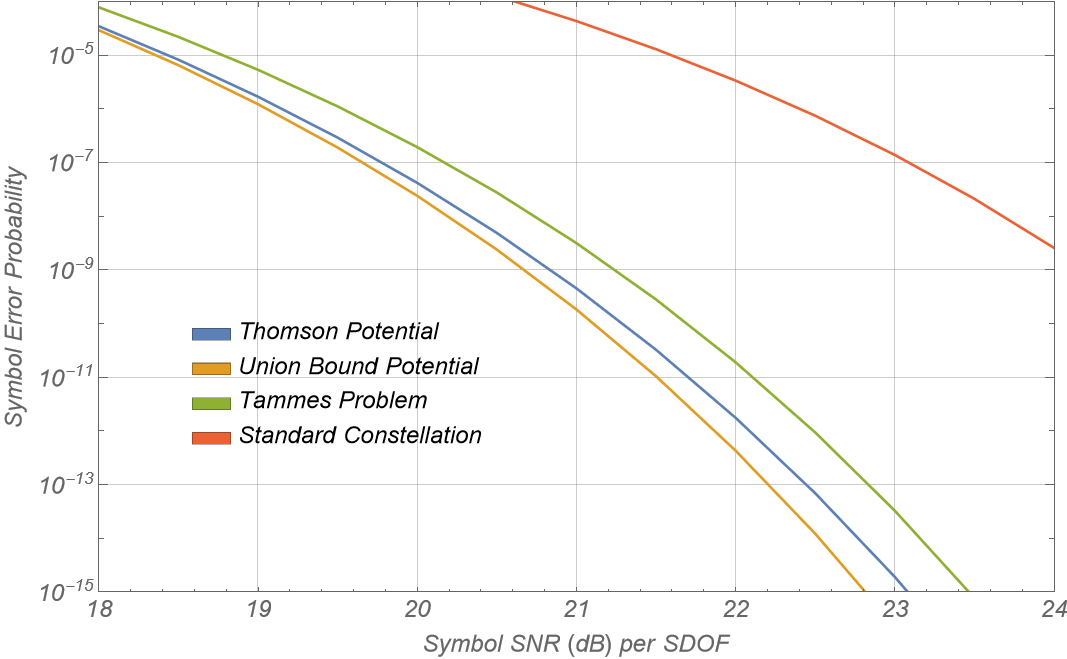}
    \caption{Performance comparison of different $(4,64)$-MVM constellations optimized using various potential functions and algorithms.}
    \label{fig:potential comparison}
\end{figure}

We use the union bound of Corollary \ref{cor:UBsumExact} to compare the performance of various $(4,64)$-MVM constellations obtained via different optimization methods. Fig. \ref{fig:potential comparison} shows the symbol error probability $\Pe$\nomenclature[$pes$]{$\Pe$}{Symbol Error Probability} as a function of the symbol SNR per SDOF. The blue and orange curves correspond to constellations obtained using the gradient descent method with a Thomson (Coulomb) potential and with the union bound based on Corollary \ref{cor:UBsumExact} as an objective function, respectively. The green curve is a numerical approximation of a solution to the Tammes problem using the Matlab code provided by \cite{jasper2019game}. Finally, as a baseline for our analysis, the red curve corresponds to a standard Jones hypercube constellation (cf. Sec. \ref{subsection: numerical details}).

Given the different algorithmic approaches and computational complexities of these methods, the parameters are selected in such a way that each implementation takes roughly the same amount of computing time in order to provide a fair comparison. Using the union bound as the objective function yields the best performing constellation, as befits its intrinsic nature, despite its high computational complexity resulting in fewer gradient descent steps in the allotted time. Belying its extrinsic motivation, the Thomson method performs remarkably well, with only a slightest penalty compared to the Union Bound potential. The Tammes Problem method also performs quite well, with only a marginal performance loss compared to the Union Bound method.  Finally, we observe that all three numerical optimizations outperform the standard Jones hypercube constellation by nearly 3 dB. 

\subsection{Simplex MVM constellations}
\begin{figure}[!htb]
	\centering
	\includegraphics[width=3in]{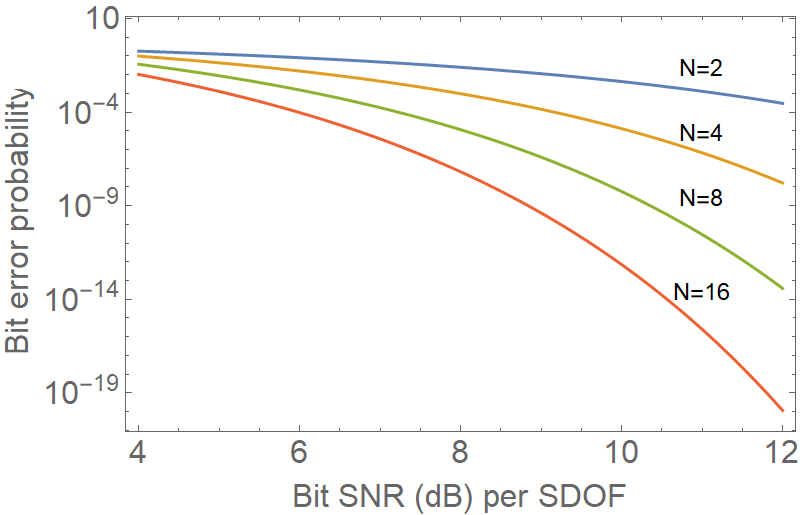}
	\caption{Bit error probability of $M-$ary MVM based on SIC-POVM vectors vs the bit SNR per spatial degree of freedom.}
	\label{fig:MVMperformance}
\end{figure}

 In Fig. \ref{fig:MVMperformance}, we plot the upper limit of the bit error probability, given by the union bound, for an optically-preamplified SIC-POVM  MVM DD receiver with matched optical filters, as a function of the electronic bit signal-to-noise ratio (SNR) per spatial degree of freedom. The Jones space dimension varies in the interval $N=2$--$16$ in power-of-two increments for different lines from top to bottom. The accuracy of the curves has also been checked by Monte Carlo simulation and the numerical data agree asymptotically with the analytical curves, but the Monte Carlo simulation results have been omitted from Fig. \ref{fig:MVMperformance} to avoid clutter. We observe that the  bit SNR required to achieve a given bit error probability decreases as $N$ increases, since the Euclidean distance between two SIC-POVM Stokes vectors increases as $\mathrm{d}=\sqrt{{2N^{2} }/({N^{2} -1}) }$. 
 
\subsection{Performance comparison of various modulation formats}
Armed with Corollary \ref{cor:UBsumExact}, we want to compare the performance of MVM with that of conventional modulation formats for short-haul transmission and optically-preamplified direct-detection. 

For a fair comparison, we want to select the MVM constellation cardinality so that MVM exhibits the same spectral efficiency as conventional modulation formats. In single-mode transmission, spectral efficiency is defined as the ratio of the net bit rate after FEC to the channel bandwidth. Here, we use the following definition of the spectral efficiency per SDOF: Let the symbol interval be $T_s$ and the symbol rate be $R_s=T_s^{-1}$. Assuming ideal Nyquist pulses and carrier modulation, the signal bandwidth $B_s$ is equal to the symbol rate $R_s$. Suppose that the bit interval is $T_b$ and the bit rate is $R_b$. Let $M$ be the number of constellation symbols. Then, $k=\log_2 M$ bits are transmitted per symbol interval. We define the spectral efficiency per SDOF as $\eta:= {R_b}/{(N B_s)}$. Since $T_s=k T_b$ and  $R_b=k R_s$, the spectral density per spatial degree of freedom is $\eta=k N^{-1}$. 

For instance, for SIC-POVMs, there are $M=N^2$ constellation points and, therefore, the spectral density per spatial degree of freedom is $\eta=2 N^{-1} \log N $. 
Consequently, by increasing the dimensionality $N$ of Jones space, the normalized spectral efficiency per SDOF decreases.

For illustration, suppose we have an ideal homogeneous MCF with eight identical single-mode cores. The most straightforward way to use this fiber is to transmit $8$ independent parallel channels, each carrying a binary signal, e.g., based on either intensity modulation (IM), binary DPSK (DBPSK), or binary SVM (BSVM). When ideal Nyquist pulses with zero roll-off factor are used, all the aforementioned modulation formats can achieve a theoretical spectral efficiency of $0.5 \, \, \mathrm{b/s/Hz/SDOF}$.

Alternatively, rather than using the $8$ cores independently, we can transmit a single MVM channel by sending pulses over all eight cores in parallel, i.e., simultaneously utilizing all 16 available spatial degrees of freedom (SDOFs) .   Therefore, we should choose $256$ MVM, which results in a spectral  efficiency is 0.5 b/s/Hz/SDOF as well. In the $255$-dimensional generalized Stokes space, the optimal (16,256)-MVM constellation corresponds to a 256-simplex \cite{Roudas:ECOC21}.

\begin{figure}[!htb]
	\centering
	\includegraphics[width=3in]{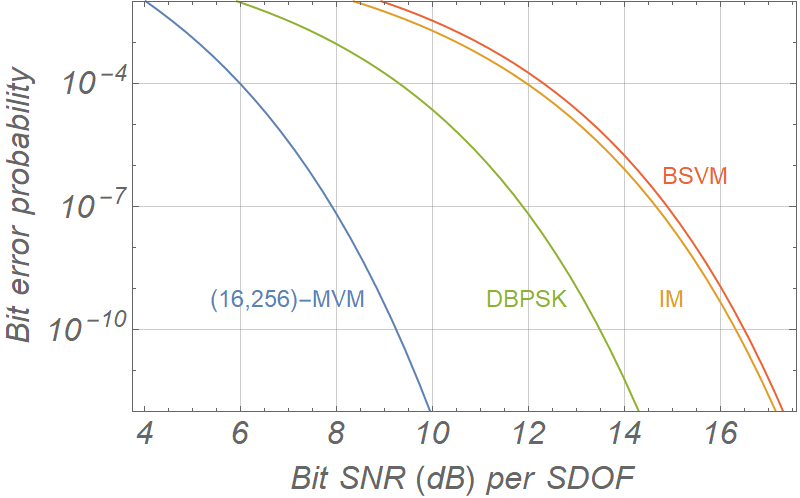}
	\caption{Bit error probability for $(16,256)$-MVM in comparison to conventional modulation formats for an $8$-core MCF.}
	\label{fig:AMFs comparison}
\end{figure}

In Fig. \ref{fig:AMFs comparison}, we present analytical plots of the bit error probability vs. the bit SNR per SDOF at the decision device. Single-polarization, optically-preamplified, direct-detection receivers require $15.83\,  \, \mathrm{dB}$, $13\,\,  \mathrm{dB}$, and $16\, \, \mathrm{dB}$ for IM \cite{humblet1991bit}, BDPSK \cite{humblet1991bit}, and BSVM \cite{Kikuchi2020}, respectively, to achieve a bit error probability of $10^{-9}$. In contrast, the $(16, 256)$-MVM optically-preamplified, direct-detection receiver requires only $8.84 \,\, \mathrm{dB}$, to achieve the same bit error probability. This corresponds to bit SNR gains of $4.16\,\, \mathrm{dB}$, $7\, \, \mathrm{dB}$, and $7.16 \, \, \mathrm{dB}$ over BDSPK, IM, and BSVM, respectively. We conclude that the use of MVM can greatly improve system performance over conventional modulation formats at the expense of transceiver complexity \cite{Roudas:ECOC21}.

\subsection{MVM performance for various $(N,M)$ pairs}
In this subsection, we present geometrically-optimized  signal sets  that correspond to the densest sphere packing in the generalized Stokes space. We show that the best trade-off between spectral and energy efficiency occurs for simplex constellations.

\begin{figure}[!htb]
	\centering
	\includegraphics[width=3in]{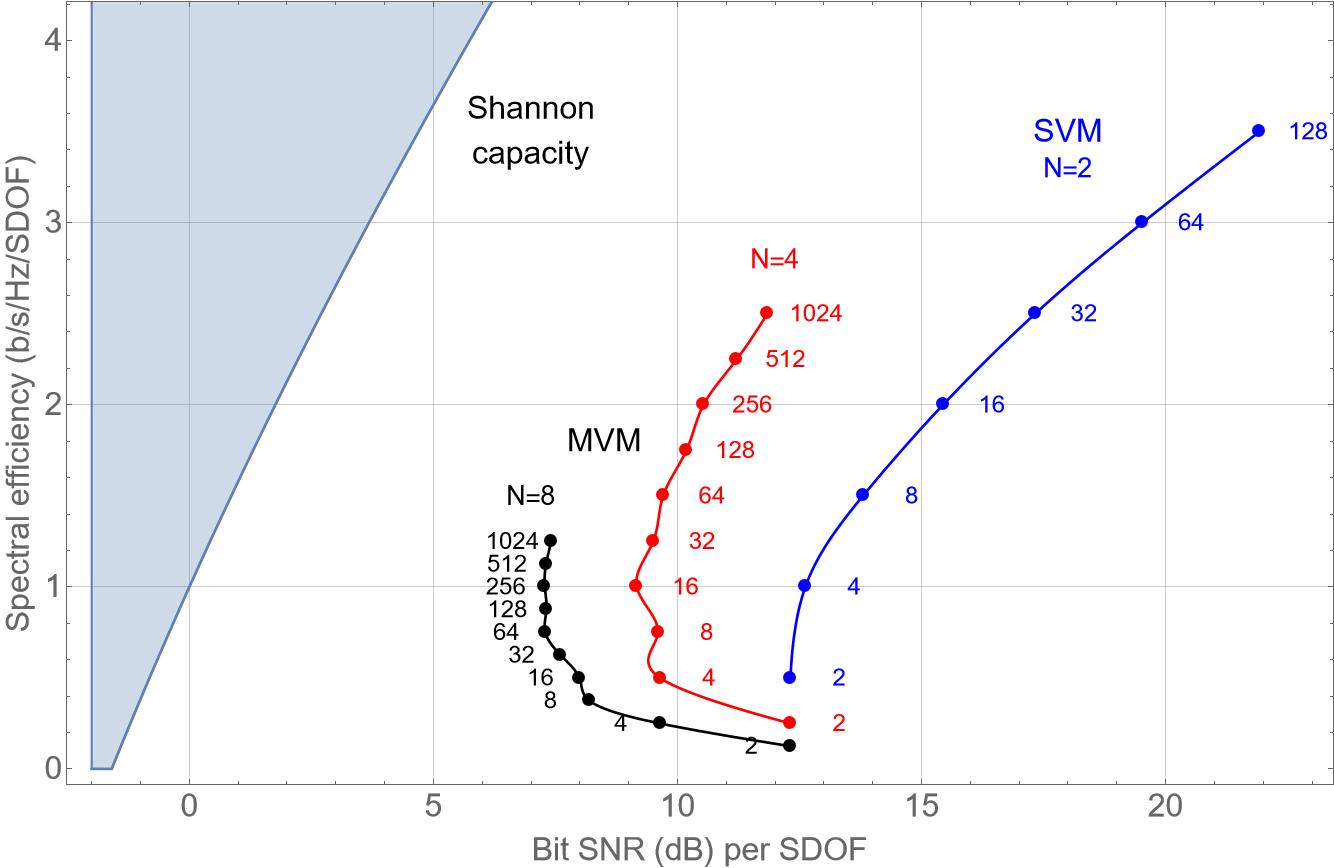}
	\caption{MVM spectral efficiencies per spatial degree of freedom (SDOF) vs the bit SNR per SDOF required to achieve a bit error probability of $10^{-4}$ for different degrees of freedom $N$ and constellation cardinalities $M$. Blue, red, and black curves correspond to $N=2,4,8$, respectively. The number listed next to each point corresponds to the constellation cardinality.}
	\label{fig:SEplot_MN_MVM}
\end{figure}

Fig. \ref{fig:SEplot_MN_MVM} shows the MVM spectral efficiency per SDOF vs the bit SNR per SDOF required to achieve a bit error probability of $10^{-4}$. Each curve corresponds to a different degree of freedom $N$, and each point within a curve corresponds to a different constellation cardinality $M$.  It is worth mentioning that these graphs represent geometrically-shaped constellations with optimized bit-to-symbol mapping. Non-optimized constellations lie on the right of these graphs. 
Furthermore, the vertices of different graphs in Fig. \ref{fig:SEplot_MN_MVM}  correspond to simplex constellations. Interestingly, the best combination of  spectral efficiency per SDOF and receiver sensitivity is achieved for SIC-POVMs. Higher spectral efficiencies can be obtained with a modest bit SNR penalty by switching to a constellation with more points, especially in higher-dimensional settings.

For the qualitative interpretation of results of Fig. \ref{fig:SEplot_MN_MVM}, we need to take a closer look at the evaluation of error probability. The leading term of the asymptotic expression for the pairwise symbol error probability based on the union bound is given by \eqref{eq:asymptoticBehavior}.

For $M>N$, the Welch–Rankin bound on $\gamma$ is written as \cite{jasper2019game}
$\gamma\geq \sqrt{{M-N}/{N(M-1)} }.$ The Welch–Rankin bound on $\gamma$ is not tight when the signal set cardinality tends to infinity. Below, we  estimate $\gamma$ from geometric arguments.

\begin{figure}[ht] 
	\centering
\captionsetup[subfigure]{justification=centering}
	\begin{subfigure}{0.5\linewidth}
		\centering
		\includegraphics[width=1.5in]{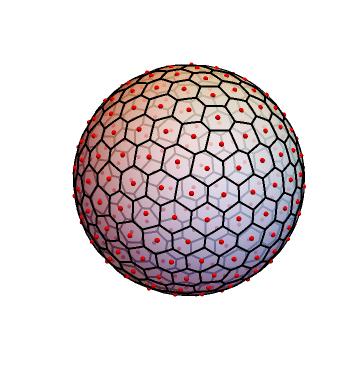}  
		\caption{} 
	\end{subfigure}
 \hfill
	\begin{subfigure}{0.5\linewidth}
		\centering
		\includegraphics[width=1.5in]{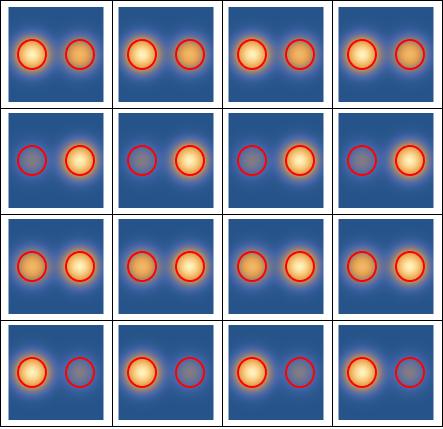}  
		\caption{} 
		\label{fig:MVM_SIC} 
  	\end{subfigure}
	\caption{(a) Optimized constellation and spherical Voronoi cells for $N=2, M=256$, obtained by solving the Thomson problem; (b) Intensity plots of the optimal MVM signal set for $N=4$, $M=16$,  over a two-core multicore fiber with identical uncoupled single-mode cores.}
	\label{fig:fig17Aux}
\end{figure}

Fig. \ref{fig:fig17Aux} shows the optimal Thomson constellation and the partitioning of the sphere into Dirichlet (Voronoi) cells for $N=2, M=256$. In general, for $N=2$ and for large $M$’s, the Dirichlet cells for an optimal configuration are mostly hexagonal \cite{Saff}. For simplicity, let's assume that the constellation points form an ideal hexagonal lattice. The Dirichlet cell for a two-dimensional hexagonal lattice is a regular hexagon of side $d/\sqrt{3}$, where $d$ is the minimum Euclidean distance between pairs of points.  The area of each cell is $\delta A=\sqrt{3}d^2/2$. We can estimate $d$ if we divide the area of the unit sphere $S^2$, equal to $A= 4\pi$, by the total area of $M$ cells. We obtain the estimate
$d^2 \approx { {8\pi}/({\sqrt{3}M})}.$ 
We observe that, in the asymptotic limit of large $M$'s, the Euclidean distance is inversely proportional to the square root of the number of points $M$. By combining the formulas \cite{Gordon2000}
$d^2=\norm{\hat{s}-\hat{s}'}^2=2(1-\hat{s}\cdot \hat{s}')$,$
    \gamma^2 =|\ip{s}{s'}|^2=(1+\hat{s}\cdot \hat{s}')/2,$
and using the first-order Taylor expansion of the square root of $\gamma^2$, we obtain the average symbol error probability
$\bar{P}_{e|s}\sim\exp\left[-{\pi\gamma_s}/({2\sqrt{3}M)}\right].$

The average bit error probability for a Gray-like bit-to-symbol mapping is related to the average symbol error probability by $\bar{P}{e|b}\simeq\bar{P}{e|s}/k$, where $k:=\log_2 M$. For quasi-orthogonal signal sets, it is related by $\bar{P}{e|b}\simeq M\bar{P}{e|s}/[2(M-1)]\simeq \bar{P}_{e|s}/2$. Gray-like bit-to-symbol mappings are expected at large constellation cardinalities $M$, while orthogonal signal sets exist for $M<N$ and quasi-orthogonal signal sets occur for $N<M<N^2$. 
In general, the difference in SNR between the different bit-to-symbol mappings is asymptotically small. For the purposes of qualitatively understanding the results shown in Fig. \ref{fig:SEplot_MN_MVM}, it is reasonable to assume that $\bar{P}_{e|b}\simeq\bar{P}_{e|s}$.

The spectral efficiency per SDOF for MVM is defined as $\eta:=k/N$ and the symbol SNR per SDOF is related to the bit SNR per SDOF via $\gamma_s:=k \gamma_b$.
 For a given average bit error probability, we can write for $N=2$ (SVM case) that
\begin{align}
\eta\sim \frac{\gamma_b \mathrm{(dB)}}{20\log{2}} .
\label{eq:asymptoticBehaviorN2}
\end{align}
Using a similar geometric argument for $N>2$ (MVM case), we find that $d^2\sim M^{-\frac{1}{N-1}}$ and we can write
\begin{align}
	\eta\sim \frac{N-1}{N}\frac{\gamma_b \mathrm{(dB)}}{10\log{2}}
	\label{eq:asymptoticBehaviorNlt2}
\end{align}
for a given average bit error probability.

Rephrasing the above expressions, we expect that the slope $\eta/{\gamma_b (dB)} \sim 0.16$  for $N=2$ at large constellation cardinalities $M$ and will increase towards 0.33 as $N\rightarrow{\infty}$, which is approximately the slope of the Shannon capacity formula. 

At the opposite extreme, $\gamma= 0$ for orthogonal signal sets  $M<N$ and we expect that
\begin{align}
	\eta\sim 10^{-\gamma_b \mathrm{(dB)} / 10}
	\label{eq:asymptoticBehaviorQO}
\end{align}
for a given average bit error probability. 


Using the preceding asymptotic analysis, we consider the results shown in Fig. \ref{fig:SEplot_MN_MVM}. The MVM spectral efficiency $\eta$ is generally expected to follow a C-shaped curve when plotted against the bit SNR per SDOF $\gamma_b\, \mathrm{(db)}$. The upper part of the curve will increase linearly with the bit SNR per \eqref{eq:asymptoticBehaviorN2} and \eqref{eq:asymptoticBehaviorNlt2}, while the lower part of the curve will decrease exponentially with the bit SNR per \eqref{eq:asymptoticBehaviorQO}. Each curve's apex occurs for simplex constellations with $M=N^2$, where $\gamma^2=(N+1)^{-1}$. We thus conclude that simplex constellations are best balance between energy and spectral efficiency for $N>2$. An example of a simplex signal set for $N=4$ and $M=16$ is shown in Fig. \ref{fig:fig17Aux}.


\subsection{Spectral efficiency vs energy efficiency trade-offs}
\begin{figure}[!htb]
	\centering
	\includegraphics[width=3in]{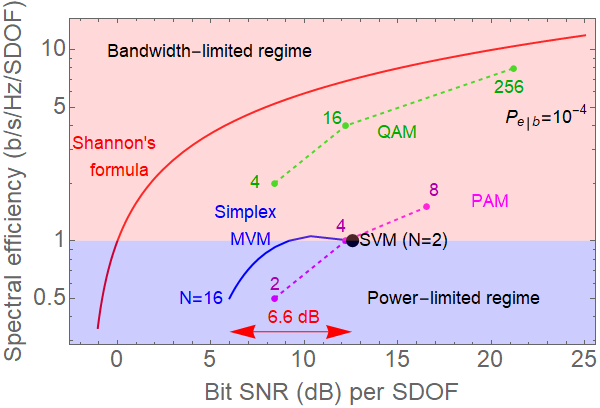}
	\caption{SIC-POVM MVM spectral efficiencies per spatial degree of freedom (SDOF) vs the bit SNR per SDOF required to achieve a bit error probability of $10^{-4}$ (in blue). Results for coherent PAM, QAM, and for SVM DD for various constellation cardinalities are also shown in magenta, green, and black, respectively.}
	\label{fig:SE}
\end{figure}

In Fig. \ref{fig:SE}, we plot  the change in spectral efficiency per SDOF for SIC-POVM MVM for different $N$ as a function of the bit SNR per SDOF required to achieve a bit error probability of $10^{-4}$ (in blue). On the same figure, we graph Shannon's formula for the spectral efficiency of an additive white Gaussian noise (AWGN) channel (in red) \cite{Proakis}. 
The maximum spectral efficiency for simplex MVM is equal to $1.06$ b/s/Hz/SDOF and occurs for $N = 3$. Similarly, the spectral efficiency for $N = 2$ and $N=4$ is 1 b/s/Hz/SDOF. This means that, at best, the spectral density of the simplex MVM is approximately equal to that of binary intensity modulation per SDOF for low $N$’s and decreases thereafter with increasing $N$. 

Notice that simplex MVM DD over SDM fibers offers 6.6 dB sensitivity improvement compared to conventional simplex SVM over SMFs ($N=2$), at the expense of spectral efficiency per SDOF. 

Based on our analysis, we conclude that using MVM DD over SDM fibers could potentially be beneficial, since the spatial degrees of freedom in SDM fibers are utilized as one channel instead of as individual channels, as is standard engineering practice. In comparison to SVM DD over SMFs, MVM offers a greater degree of flexibility for balancing energy consumption and spectral efficiency.

\section{Conclusion}
Renewed interest in direct-detection systems  is based on the assumption that the cost and energy consumption of coherent receivers designed for long-haul transmission will be prohibitive  for short-reach optical interconnects in the near and medium term \cite{Perin2018}, \cite{PerinJLT21}.

 Adopting the optimistic view that the price of SVM transceivers  will eventually become affordable  through photonic integration before that of coherent receivers, we went one step further and envisioned the use of SVM-like spatial modulations (referred to as {\em MVM}) over  multimode and multicore  fibers or free-space.

In this paper, we  investigated the merits of MVM with equipower signal sets, which is a direct extension of PolSK for the generalized Stokes space. In other words, we limited ourselves to a subset of the full spatial modulation/direct-detection family set. We derived an analytical upper limit for the back-to-back performance of $M$-ary MVM over $N$ spatial degrees of freedom in the amplified spontaneous emission (ASE) noise-limited regime. 

We also elaborated on the following topics: (i) The optimal MVM transceiver architecture; (ii) The use of simplex MVM constellations based on symmetric, informationally complete, positive operator valued measure (SIC-POVM) vectors; (iii) The design of M-ary geometrically-shaped constellations obtained by numerical optimization  of  various  objective  functions  using  the  method of  gradient  descent; and (iv)  The  optimal  bit-to-symbol  mapping using simulated annealing. 

We showed that it is potentially beneficial to use  MVM DD over SDM fibers, i.e., to use spatial degrees of freedom in SDM fibers together as a single channel instead of individually as separate channels, per standard engineering practice. Compared to SVM DD over SMFs, MVM DD over SDM fibers offers greater flexibility for better trade-offs between energy consumption and spectral efficiency.

 The successful  commercialization  of MVM eventually depends on technoeconomics. MVM, like other advanced direct-detection techniques for spectrally-efficient transmission \cite{Kikuchi2014}, \cite{mecozzi2016kramers}, \cite{yoshida_JLT_19}, \cite{Chen_JLT_20}, requires several parallel optical branches followed by ADCs and DSP, all of which increase cost and energy consumption in comparison to $M$-ary PAM and approach or even exceed the complexity of coherent receivers. Therefore, MVM's future commercial viability depends on the development of inexpensive  silicon photonic (SiP) integrated circuits and application-specific integrated circuits (ASICs) for DSP. In the end, it is likely that low-power, "light"  coherent receivers that avoid high-speed  analog-to-digital converters (ADCs) and digital signal processing (DSP) might eventually prevail against all direct-detection alternatives in intra- and inter-data center links \cite{Perin2018}, \cite{PerinJLT21}, \cite{zhou_beyond_2020}, \cite{morsy-osman_dsp-free_2018}, \cite{jia_coherent_2021}, \cite{rizzelli_scaling_2021}.

\appendices



\section{Pairwise symbol error probability}\label{app:mainProofPes}
\noindent The goal of this appendix is to  compute  the  pairwise error probability  $\Peb^{m' | m}$   defined by \eqref{eq:mPrimeUnionBoundBis} and prove \eqref{mainExact:eq}.

\subsection{Geometric Setup}

\noindent 
We fix $m \neq m'$ and,  to simplify expressions, set $\ket{s}:=\ket{s_m}$ and $\ket{s'}:=\ket{s_{m'}}$. 
We assume that
\begin{subequations}
	\begin{align}
	\ip{s} &=\ip{s'}=1, \\ \gamma & :=\ip{s}{s'}>0.
	\label{eq:decision_inequality}
\end{align}
\end{subequations}
In \eqref{eq:decision_inequality},  we dropped the absolute value on $\ip{s}{s'}$ because, since we deal with two vectors in isolation and only the projection operators ${\bf S}=\dyad{s}$ and ${\bf S}'=\dyad{s'}$ matter,  we can adjust the phase of $\ket{s'}$ so that  $\ip{s'}{s}$ is a positive real.
As the first step, we reduce the considerations to the two-dimensional complex subspace $\Sigma$ spanned by $\ket{s}$ and $\ket{s'}$ and derive analytical expressions in a convenient orthonormal basis for $\Sigma$. 

We  express $\Peb^{m' | m}$ as 
\begin{align}
\label{PebQuad:eq}
	\Peb^{m' | m} &= \Prob\left(  |\ip{r}{s'}|^2 -  |\ip{r}{s}|^2 \geq 0 \right) \notag\\
	&= \Prob\left( \mel{r}{{\bf {\bf \Delta}}}{r} \geq 0 \right),
\end{align}
where we introduced the difference of dyads \nomenclature[$delta$]{${\bf \Delta}$}{Difference of dyads, ${\bf \Delta}:= \ket{s'}\bra{s'}-\ket{s}\bra{s}$}
${\bf \Delta} := {\bf S'}-{\bf S} $ and the associated quadratic form\footnote{Note that this formulation makes it clear that ${\mathcal D}^{m' | m}$ is bounded by $3D$-cone in $\Sigma$ treated as a 4D real space.}   
\begin{align}
	|\ip{r}{s'}|^2 - |\ip{r}{s}|^2 &= \trace\left( {\bf R}{\bf S}'\right) - \trace\left( {\bf R}{\bf S}\right) \notag \\
	&= \trace\left( {\bf R}{\bf \Delta} \right) 
	= \mel{r}{{\bf \Delta}}{r}. 
\end{align}




\noindent The following three real \emph{length parameters} will play a key role: 
\begin{subequations}
\begin{align}
	\label{deltarhodef:eq}
	\delta &:= \sqrt{1-\gamma^2},  \\
	\rho_\pm &:= \sqrt{\frac{1 \pm \delta}{2}}.
	\label{deltarhodef:eq2}
\end{align}
\end{subequations}

For ease of reference we record that 
\begin{subequations}
\begin{align}
	\label{prels:eq}
	\gamma &= \sqrt{1-\delta^2} = 2\rho_+ \rho_-  \\
	\rho_\pm - \rho_\mp \gamma &= (1-2\rho_\mp^2) \rho_\pm  = \pm \delta \rho_\pm. \label{prels2:eq}
\end{align}
\end{subequations}
We also introduce two  vectors in $\Sigma$, $\ket{u_+}$ and $\ket{u_-}$,   defined as
\begin{align}
	\label{eq:ONbasis}
	\ket{u_\pm} 
	&:=\pm \frac{\rho_\pm \ket{s'} - \rho_\mp \ket{s} }{\delta}.
\end{align}
Their scalar components along $\ket{s'}$ and $\ket{s}$ are found, via \eqref{prels:eq} and \eqref{prels2:eq}, to be
\begin{subequations}
\begin{align}
	\label{eq:inProdS}
	\ip{s'}{u_\pm} &=\pm \frac{\rho_\pm - \rho_\mp\gamma}{\delta  } = \rho_\pm  \\ 
	\ip{s}{u_\pm}& =\pm \frac{  \rho_\pm\gamma - \rho_\mp}{\delta  } =  \rho_\mp.
		\label{eq:inProdS2}
\end{align}
\end{subequations}
Computing ${\bf \Delta} \ket{u_\pm}$ as the difference 
 of the projections onto  $\ket{s'}$ and $\ket{s}$ (and then using \eqref{eq:ONbasis}) gives 
\begin{align}
	{\bf \Delta} \ket{u_\pm} 
	=   \rho_\pm\ket{s'} -  \rho_\mp\ket{s} = \pm \delta \ket{u_\pm}.
\end{align}
Thus  $\ket{u_\pm}$ are eigenvectors of ${\bf \Delta}$ with eigenvalues $\pm \delta$, respectively. 
They are normalized since  combining  \eqref{eq:ONbasis} and \eqref{eq:inProdS} yields
They are normalized since  combining  \eqref{eq:ONbasis} and \eqref{eq:inProdS}, \eqref{eq:inProdS2} yields
\begin{align}
	\ip{u_\pm} &= \pm \frac{1}{\delta} \ip{ \ (\rho_\pm \ket{s'} - \rho_\mp \ket{s}) \ }{u_\pm} \notag \\
	&= \pm \frac{1}{\delta} \left( \rho_\pm^2 -  \rho_\mp^2 \right) 
	=1. 
\end{align}

Because ${\bf \Delta}$ is Hermitian  of  rank two, the eigenvectors $\ket{u_\pm}$ are orthogonal and the remaining eigenvalue of ${\bf \Delta}$, other than $\pm \delta$, is zero (with the orthogonal complement $\Sigma^\perp$ as its eigenspace).   
The underlying geometry is simple: Examining \eqref{eq:ONbasis}, we see that  $\ket{u_\pm}$ sit in the real sub-plane inside $\Sigma$  spanned by $\ket{s'}$ and $\ket{s}$.
The vectors  $\ket{s'}$ and $\ket{s}$ form an acute angle  (by virtue of our initial phase rotation).
From \eqref{eq:inProdS} and \eqref{eq:inProdS2},   $\ip{s'}{u_+}= \rho_+ = \ip{s}{u_-}$,  so this acute angle  is positioned symmetrically within the right angle formed by  $\ket{u_\pm}$. One could say that $\ket{u_\pm}$ are the result of \emph{symmetrically opening up} $\ket{s'}$ and $\ket{s}$ to be orthogonal.






\subsection{Signal Decomposition}


\noindent With our orthonormal basis $\ket{u_\pm}$ of $\Sigma$ in hand, we  
orthogonally decompose the noise
\begin{align}
	\label{noisedecomp:eq}
	\ket{n} = \ket{n_+} + \ket{n_-} + \ket{\tilde{n}},
\end{align}
where  the components along  $\ket{u_\pm}$ are 
\begin{equation}
	\label{eq:ONbasisTwoDim}
	\ket{n_+} := \ip{n}{u_+} \ket{u_+} \quad \text{and} \quad
	\ket{n_-} := \ip{n}{u_-} \ket{u_-},
\end{equation}
and $\ket{\tilde{n}}$ is the component orthogonal to $\Sigma$.
(Going forward, tilde indicates components orthogonal to $\Sigma$.)
Inverting \eqref{eq:ONbasis}, we get the analogous decomposition 
of the symbols
\begin{subequations}
\begin{align}
	\label{symbDecomp:eq}
	\ket{s} &=  \rho_- \ket{u_+} + \rho_+ \ket{u_-}, \\
	\ket{s'} &=  \rho_+ \ket{u_+} + \rho_- \ket{u_-}.
	\label{symbDecomp:eq2}
\end{align}
\end{subequations}

Because the Gaussian noise $\ket{n}$ is symmetric with respect to phase rotations, we can disregard the random phase in \eqref{eq:AWGNJoneschannelBis} and express the Jones vector representing the incoherently received signal as $\ket{r} = \ket{s} + \ket{n}$. 
Putting together \eqref{noisedecomp:eq} and \eqref{symbDecomp:eq}, \eqref{symbDecomp:eq2}, reveals its components along  $\ket{u_\pm}$ as 
equal to 
\begin{subequations}
\label{rcomp:eq}
\begin{align}
	\ket{r_+} &= \rho_-\ket{u_+} + \ket{n_+},
	\\ 
	\ket{r_-} &= \rho_+\ket{u_-} + \ket{n_-},
\end{align}
\end{subequations}
with the squared magnitudes consequently given by 
\begin{equation}
\label{rpmSize:eq}
  \ip{r_\pm}  = \rho_{\mp}^2 + 2 \rho_{\mp} \Re \ip{n_\pm}{u_\pm} + \ip{n_\pm}.
\end{equation}

The full $\ket{r}$ decomposes into orthogonal components,
\begin{equation}
	\ket{r} = \ket{s} + \ket{n} 
	= \ket{r_+} + \ket{r_-} + \ket{\tilde{r}},
\end{equation}
along the eigenspaces of ${\bf \Delta}$ for eigenvalues $\delta$, $-\delta$, and $0$, respectively. 

Using  ${\bf \Delta} \ket{\tilde{r}}=0$ as well as ${\bf \Delta} \ket{r_\pm}= \pm \delta \ket{r_\pm}$ and  $\ip{r_-}{r_+}=0$,            
 the quadratic form simplifies to  
\begin{align}\label{pmDeltaProd:eq}
	\mel{r}{{\bf \Delta}}{r}
	&= \mel{\ \ket{r_+} + \ket{r_-} + \ket{\tilde{r}} \ }{\ {\bf \Delta} \ }{ \  \ket{r_+} + \ket{r_-} + \ket{\tilde{r}} \ } \notag\\
	&=  \mel{r_+}{{\bf \Delta}}{r_+}  + \mel{r_-}{\bf \Delta}{r_-} \notag\\
		&= \delta \ip{r_+} - \delta \ip{r_-}. 
\end{align}
Finally, substituting \eqref{rpmSize:eq}, yields 
\begin{align}
	\label{mainMelQuadBis:eq}
	 \mel{r}{{\bf \Delta}}{r} 
	&= \delta \bigg[\rho_-^2 -\rho_+^2 
	+ 2 \Re \left\{ \rho_-  \ip{u_+}{n_+}  - \rho_+ \ip{u_-}{n_-} \right\} \notag\\
    &\qquad +\ip{n_+} - \ip{n_-}\bigg].
\end{align}


\subsection{Pairwise symbol error probability calculation}
\noindent We are ready to derive the closed form \eqref{mainExact:eq} for  $\Peb^{m' | m}$ by identifying the relevant probability distributions associated to the quadratic form.  
We can describe the points of $\Sigma$ by their components $x_-+\iota y_-$ and $x_++\iota y_+$ with respect to the orthonormal basis $\ket{u_\mp}$. 
Accordingly, we have four independent real Gaussian random variables with variance $\nvar$:
\begin{align}
    x_\mp := \Re \ip{n}{u_\mp} \quad \text{ and } \quad 
    y_\mp := \Im \ip{n}{u_\mp}. 
\end{align}
The last equation of the previous section, \eqref{mainMelQuadBis:eq}, reads
\begin{align}
  \frac{1}{\delta} \mel{r}{{\bf \Delta}}{r}
  &= \rho_-^2 -\rho_+^2 +  2 \rho_-x_+ - 2 \rho_+ x_- \notag\\
     &\quad +x_+^2 + y_+^2 - x_-^2 -y_-^2.
\end{align}
So, upon completing the squares,  the sought pairwise error probability in \eqref{PebQuad:eq}  
is 
\begin{align}
    \Pbin
    &= \Prob\left( \mel{r}{{\bf \Delta}}{r} \geq 0 \right) \notag\\
    &= \Prob\left(\left( \rho_- + x_+ \right)^2 + y_+^2 \geq 
    \left(\rho_+ + x_- \right)^2 +  y_-^2  \right).
\end{align}
This is to say that  
\begin{align}
  \label{PebViariceIneqProb:eq}
  \Pbin = \Prob\left( \psi_- \geq \psi_+  \right),
\end{align}
 where we introduced two independent Rice-distributed random variables  
\begin{subequations}
\begin{align}
  \psi_- &:= \sqrt{\left(x_+ +\rho_-\right)^2 + y_+^2},  \\
  \psi_+ &:= \sqrt{\left(x_- +\rho_+\right)^2 + y_-^2}
\end{align}
\end{subequations}
with  reference distances $\rho_-$ and  $\rho_+$, respectively,   and a common scale parameter $\sigma$. 
The PDFs of $\psi_\pm$ are 
\begin{align}
  \label{eq:RiceCDFbis}
 f_\pm(x) = \frac{x}{\nvar}
    \exp\left({-\frac{x^2+\rho_\pm^2}{2\nvar}}\right)
    I_0\left( \frac{x \rho_\pm}{\nvar} \right) ,
\end{align}
 with the corresponding tail (complementary) distribution functions \cite{Proakis}
\begin{equation}
   \label{eq:nonCentchiSquaredCDF}
   \Prob(\psi_\pm \geq x) = Q_{1}\left( \frac{\rho_\pm}{\sqrt{\nvar}},\frac{x}{\sqrt{\nvar}} \right).
 \end{equation}
 Recall that $Q_1$  stands for the \emph{Marcum function} defined by \eqref{eq:MarcumDef}.

 Thus, formula \eqref{PebViariceIneqProb:eq} can be represented by a single integral:
 \begin{align}
  \label{singleInRep:eq}
  &\quad \Pbin \notag\\
  &= \int_0^\infty \Prob\left( \psi_- \geq x  \right)\, f_+(x) dx \notag \\
  &= \int_0^\infty Q_{1}\left( \frac{\rho_-}{\sqrt{\nvar}},\frac{x}{\sqrt{\nvar}} \right)
    \frac{x}{\nvar} 
    \exp\left({-\frac{x^2+\rho_+^2}{2\nvar}}\right) 
    I_0\left( \frac{x \rho_+}{\nvar} \right) \,
    dx. 
 \end{align}
 The integral can be computed
 and is given by a formula from \cite{Benedetto1987}
 to the effect that 
 \begin{align}
    \Pbin
    &= \Prob(\psi_- \geq \psi_+) \notag\\
    &= Q_1(\aa,\bb)-\frac{1}{2}
    \exp\left({-\frac{\aa^2+\bb^2}{2}}\right)
    I_0(\aa \bb),
    \label{eq:proof_ub}
 \end{align}
 where  (recalling  $\gamma_s = \frac{1}{2\nvar}$, per \eqref{gammasSigma:eq})
 \begin{subequations}
 \begin{align}
   \aa  &:=\frac{\rho_-}{\sqrt{2 \nvar}}=\rho_-\sqrt{\gamma_s}, \\
   \quad  \bb&:=\frac{\rho_+}{\sqrt{2 \nvar}}=\rho_+\sqrt{\gamma_s}.
 \end{align}
\end{subequations}

 Using $\rho^2_- + \rho^2_+ = 1$ and $\rho^2_- \rho^2_+ = \frac{1-\delta^2}{4}=\frac{\gamma^2}{4}$ from \eqref{deltarhodef:eq} through \eqref{prels2:eq} gives
 \begin{subequations}
 \begin{align}
   \label{aabbRels:eq}
   \aa^2 + \bb^2 &=(\rho^2_-+\rho^2_+)\gamma_s = \gamma_s, \\
    2\aa\bb &=2\rho_-\rho_+\gamma_s = \gamma \gamma_s. 
 \end{align}
\end{subequations}
Thus \eqref{eq:proof_ub} coincides with the promised formula \eqref{mainExact:eq}.




\section{Asymptotics for large SNRs}\label{app:Asymptotics}
\noindent We derive the approximate formulas for the error probability stated in Corollary~\ref{asympt:cor} and valid for large SNR parameters $\gamma_s$, as well as the simplified approximation \eqref{mainAsymptOld:eq}. 

The exact formula \eqref{mainExact:eq} reads 
   \begin{equation}
     \label{mainExact:eqBis}
  \Pbin = Q_1\left( \rho_-\sqrt{\gamma_s},\rho_+\sqrt{\gamma_s} \right)
     -\frac{1}{2} \exp\left({-\frac{\gamma_s}{2}}\right) I_0\left(\frac{\gamma \gamma_s}{2}\right).
   \end{equation}

   For large $x:=\frac{\gamma \gamma_s}{2}$, 
    taking the first $n+1$ terms of the Hankel asymptotics given by 
   \cite{benedetto2006principles} 
    \begin{align*}
  I_0(x) &= \frac{e^x}{\sqrt{2\pi x}}
           \left(1 + \frac{1}{8x} + \frac{1 \cdot 9}{2!(8x)^2} + \frac{1 \cdot 9 \cdot 25}{3!(8x)^3}  + \ldots \right),   
   \end{align*}
   yields an approximation to the Bessel term in \eqref{mainExact:eqBis}: 
   \begin{align}
     \label{BesselAsympt:eq}
  \exp\left({-\frac{\gamma_s}{2}}\right) I_0\left(\frac{\gamma \gamma_s}{2}\right)   \sim \Bc_n    \exp \left( {-\frac{\gamma_s(1-\gamma)}{2}}\right),
\end{align}
with 
\begin{align}
     \label{BesselAsymptCoef:eq}
  \Bc_n:= \frac{1}{\sqrt{\pi}}
  \left( \frac{1}{(\gamma \gamma_s)^{\frac{1}{2}}} + \frac{\frac{1}{4}}{(\gamma \gamma_s)^{\frac{3}{2}}} + \frac{\frac{9}{32}}{(\gamma \gamma_s)^{\frac{5}{2}}} + \ldots \right)   
\end{align}
where the sum is terminated on  $\frac{(2n-1)!!^2}{4^{n}n!}(\gamma \gamma_s)^{-n-1/2}$ for $n \geq 1$.

A similar asymptotic expansion for the Marcum term $Q_1\left( \rho_-\sqrt{\gamma_s},\rho_+\sqrt{\gamma_s} \right)$ in \eqref{mainExact:eqBis} is more subtle and can be extracted from \cite{GilSeguraTemme2014} in the  form of  a linear combination of the exponential  $\exp\left({-\frac{\gamma_s(1-\gamma)}{2}}\right)$ and the error function $\erfc\left(\frac{\sqrt{\gamma_s}\sqrt{1-\gamma}}{\sqrt{2}} \right)$ with the coefficients described below. (Here $\erfc(x):=\frac{2}{\sqrt{\pi}} \int_x^\infty e^{-t^2}\, dt$.)  



To start, define $\Ce_n$ and $\Cf_n$ recursively: Let $\Ce_0:=0$ and follow with    
\begin{equation}
  \label{Ce:eq}
  \Ce_n:= \frac{1}{\frac{1}{2}-n}\left( \frac{1-\gamma}{\gamma} \Ce_{n-1} - \left(\frac{\gamma \gamma_s}{2}\right)^{\frac{1}{2}-n} \right) \quad(n\geq 1). 
\end{equation}
 Let $ \Cf_0 := \sqrt{\pi}\sqrt{\frac{\gamma}{1-\gamma}}$ and follow with 
\begin{equation}\label{Cf:eq}
  \Cf_n := \frac{1}{\frac{1}{2}-n} \frac{1-\gamma}{\gamma} \Cf_{n-1} \quad (n\geq 1).
\end{equation}
Then, using constants 
\begin{equation}\label{Acoef:eq}
  A_{n,m}:=\frac{1}{n!2^n}\frac{\Gamma(\frac{1}{2}+m+n)}{\Gamma(\frac{1}{2}+m-n)}
  = \frac{1}{n!2^n} \prod_{i=-n}^{n-1}\left(m+i+\frac{1}{2}\right),  
\end{equation}
define a multiplier 
\begin{align}
  \lambda_n:=\frac{(-1)^n}{2 \sqrt{2\pi}}\left( \frac{\rho_+}{\rho_-}  A_{n,0}- A_{n,1} \right),
\end{align}
and set
\begin{subequations}
\begin{align}
  \Ce'_n &:=\lambda_n \Ce_n, \\ \Cf'_n &:=\lambda_n \Cf_n, \\ \Ce''_n &:=\sum_{i=0}^n \Ce'_i, \\  \Cf''_n &:=\sum_{i=0}^n \Cf'_i.
\end{align} 
\end{subequations}
We note that $\Cf''_n$ only depends on $\gamma$ while $\Ce''_n$ is a linear combination of the powers 
$(\gamma\gamma_s)^{-\frac{1}{2}}$, $(\gamma\gamma_s)^{-\frac{3}{2}}$, $\ldots$, $(\gamma\gamma_s)^{-\frac{2n-1}{2}}$ with $\gamma$-dependent coefficients  (for $n \geq 1$).

The approximation given by formula (37) in \cite{GilSeguraTemme2014} reads then
\begin{align}
  \label{MarcumPerGil:eq}
    &\quad Q_1\left( \rho_-\sqrt{\gamma_s},\rho_+\sqrt{\gamma_s} \right) \notag\\
    &\sim \Ce''_n \exp\left({-\frac{\gamma_s(1-\gamma)}{2}}\right) + \Cf''_n \erfc\left(\frac{\sqrt{\gamma_s}\sqrt{1-\gamma}}{\sqrt{2}} \right). 
\end{align}

An important feature of \eqref{MarcumPerGil:eq} is that it is valid
uniformly across $\gamma \in (0,1)$ (as long as $\gamma \gamma_s$ is sufficiently large).
In \cite{Temme1986} 
 explicit error bounds are discussed together with suitable expansion termination criteria. For our purposes, using $n = 1$ gives excellent results.

To approximate the error probability $\Pbin$, as given by \eqref{mainExact:eqBis}, we combine the Marcum and Bessel approximations, \eqref{MarcumPerGil:eq} and \eqref{BesselAsympt:eq}, and obtain:
\begin{align}
  \label{ultimateApprox:eq}
     \Pbin  
  &\sim   \Cf''_n \erfc\left(\frac{\sqrt{\gamma_s}\sqrt{1-\gamma}}{\sqrt{2}} \right) \notag\\
  &\quad + \left(\Ce''_n -\frac{1}{2}\Bc_{n} \right) \exp\left({-\frac{\gamma_s(1-\gamma)}{2}}\right).
\end{align}
Corollary~\ref{asympt:cor} will follow now by using  $n=0,1$ in  \eqref{ultimateApprox:eq}. 



To streamline formulas we reach back to \eqref{deltarhodef:eq}, \eqref{deltarhodef:eq2}, and  note
\begin{equation}
  \label{rhoRatio2:eq}
  \frac{\rho_+}{\rho_-}
  =\frac{\sqrt{1+\delta}}{\sqrt{1-\delta}}
 =\frac{1+\delta}{\sqrt{1-\delta^2}}=\frac{1+\delta}{\gamma}.
\end{equation}
Also,  squaring as follows 
\begin{equation}
    (\rho_+ \pm \rho_-)^2 = \left( \sqrt{\frac{1+\delta}{2}} \pm \sqrt{\frac{1-\delta}{2}}\right)^2 = 1 \pm \sqrt{1-\delta^2},
\end{equation}
 gives \begin{equation}
  \label{rhoPMaux:eq}
   \rho_+ \pm \rho_- = \sqrt{1 \pm \gamma}.
 \end{equation}
In particular, 
\begin{equation}
  \label{rhoRatio1:eq}
   \frac{\rho_+}{\rho_-}-1=\frac{\rho_+-\rho_-}{\rho_-}= \sqrt{2}\sqrt{\frac{1 - \gamma}{1-\delta}}.
 \end{equation}


Taking $n=0$, we find  
$A_{0,0}=1$ and $A_{0,1}=1$, so 
\begin{align}
  \label{CF0:eq}
  \Cf_0'' =  \Cf_0' = \lambda_0 \Cf_0  &= \frac{1}{2\sqrt{2\pi}} \left(\frac{\rho_+}{\rho_-}-1\right)\sqrt{\pi}\sqrt{\frac{\gamma}{1-\gamma}} \notag \\
    &= \frac{1}{2}\sqrt{\frac{\gamma}{1-\delta}} \ ,
\end{align}
where we used \eqref{rhoRatio1:eq}. 
 Plugging \eqref{CF0:eq} and $\Ce''_0=0$ and $\Bc_0=(\gamma \gamma_s)^{-1/2}/\sqrt{\pi}$ (from \eqref{BesselAsymptCoef:eq}) into  \eqref{ultimateApprox:eq}  reproduces \eqref{mainAsymptZero:eq}, the first formula in  Corollary~\ref{asympt:cor}:
  \begin{align}
    \label{mainAsymZeroBis:eq}
    \Pbin
    \sim  
     &\frac{1}{2} \sqrt{\frac{\gamma}{1-\delta}} \erfc\left(\frac{\sqrt{\gamma_s}\sqrt{1-\gamma}}{\sqrt{2}} \right) \notag \\
    &\quad - \frac{1}{2\sqrt{\pi \gamma \gamma_s}} \exp\left({-\frac{\gamma_s(1-\gamma)}{2}}\right).
 \end{align}



 Taking $n=1$, we find
 $A_{1,0}=-\frac{1}{8}$ and $A_{1,1}=\frac{3}{8}$, 
  so  
\begin{align}
  \Cf_1' = \Cf_1 \lambda_1 &= \frac{1}{\frac{1}{2}-1} \frac{1-\gamma}{\gamma} \sqrt{\pi}\sqrt{\frac{\gamma}{1-\gamma}}
                             \frac{(-1)^1}{2 \sqrt{2\pi}}\left( -\frac{\rho_+}{\rho_-}\frac{1}{8}- \frac{3}{8} \right)    \notag  \\
                           &=  - \sqrt{\frac{1-\gamma}{\gamma}}
                             \frac{1}{8\sqrt{2}}\left( \frac{\rho_+}{\rho_-}+ 3 \right) \notag \\
   &= - \sqrt{\frac{1-\gamma}{\gamma}}  \frac{1}{8\sqrt{2}}\left(\frac{1+\delta}{\gamma} + 3 \right)
\end{align}
where we used \eqref{rhoRatio2:eq}. 
Thus, using \eqref{CF0:eq}, we arrive with 
\begin{align}
  \label{cfff:eq}
  \Cf_1'' = \Cf_0' +  \Cf_1'
                   &=\frac{1}{2} \sqrt{\frac{\gamma}{1-\delta}}
                     - \sqrt{\frac{1-\gamma}{\gamma}}  \frac{1}{8\sqrt{2}}\left(\frac{1+\delta}{\gamma} + 3 \right).
\end{align}

Turning attention to $\Ce''_1=\Ce_1' = \Ce_1 \lambda_1$, we have 
\begin{align}
  \Ce_1''  &= \frac{1}{\frac{1}{2}-1}\left( - \left(\frac{\gamma \gamma_s}{2}\right)^{\frac{1}{2}-1} \right)\frac{(-1)^1}{2 \sqrt{2\pi}}\left( -\frac{\rho_+}{\rho_-}\frac{1}{8}- \frac{3}{8} \right)  \notag \\
    &=  \frac{1}{8 \sqrt{\pi}}\left( \frac{\rho_+}{\rho_-}+3 \right)\left(\gamma \gamma_s\right)^{-\frac{1}{2}}.
\end{align}
Fetching  $\Bc_0 $ from  
 \eqref{BesselAsymptCoef:eq} and then using \eqref{rhoRatio1:eq} gives  
\begin{align}\label{cfffBis:eq}
  \Ce_1'' - \frac{1}{2}\Bc_0 &=  \frac{1}{8 \sqrt{\pi}}\left( \frac{\rho_+}{\rho_-}+3 \right) \left(\gamma \gamma_s\right)^{-\frac{1}{2}}-
                              \frac{1}{2} \frac{1}{\sqrt{\pi}} (\gamma \gamma_s)^{-\frac{1}{2}} \notag \\
                            &=  
                              \frac{1}{8\sqrt{\pi}} \left( \frac{\rho_+}{\rho_-} - 1 \right)\left(\gamma \gamma_s\right)^{-\frac{1}{2}} 
  \notag \\
                            &= 
                              \frac{\sqrt{2}}{8\sqrt{\pi}} \sqrt{\frac{1-\gamma}{1-\delta}} \left(\gamma \gamma_s\right)^{-\frac{1}{2}}. 
\end{align}
Subtracting one more term of the Bessel expansion  \eqref{BesselAsymptCoef:eq} yields
\begin{align}\label{ceee:eq}
  \Ce_1'' - \frac{1}{2}\Bc_1  &=  
                               \frac{\sqrt{2}}{8\sqrt{\pi}} \sqrt{\frac{1-\gamma}{1-\delta}}\left(\gamma \gamma_s\right)^{-\frac{1}{2}}   - \frac{1}{8\sqrt{\pi}}\left(\gamma \gamma_s\right)^{-\frac{3}{2}}.
\end{align}


One can check now that plugging \eqref{ceee:eq} and \eqref{cfff:eq} into \eqref{ultimateApprox:eq}  reproduces  \eqref{mainAsympt:eq}, the second formula
  in  Corollary~\ref{asympt:cor}.

  \begin{rmk}
    \label{truncEffect:rmk}
   Dropping the $\left(\gamma \gamma_s\right)^{-\frac{3}{2}}$ term in the second formula  in Corollary~\ref{asympt:cor} yields 
\begin{align}
   \Pbin 
   &\sim     
             \left[ \frac{1}{2} \sqrt{\frac{\gamma}{1-\delta}}
     - \sqrt{\frac{1-\gamma}{\gamma}}  \frac{1}{8\sqrt{2}}\left(\frac{1+\delta}{\gamma} + 3 \right) \right]  \notag \\ 
    &\quad\times\erfc\left(\frac{\sqrt{\gamma_s}\sqrt{1-\gamma}}{\sqrt{2}} \right) \notag \\
    &\quad + \frac{\sqrt{2}}{8\sqrt{\pi}}\sqrt{\frac{1-\gamma}{1-\delta}}\left(\gamma \gamma_s\right)^{-\frac{1}{2}} \exp\left({-\frac{\gamma_s(1-\gamma)}{2}}\right).
 \end{align}
This is a somewhat looser approximation for very large $\gamma_s$
  but works well for moderate values of $\gamma_s$ of interest in our applications. 
 \end{rmk}


 It remains to derive the crude approximation \eqref{mainAsymptOld:eq}. When  $\sqrt{x}=\frac{\sqrt{\gamma_s}\sqrt{1-\gamma}}{\sqrt{2}}$ is large, which happens for large $\gamma_s$ when $\gamma$ is not too close to $1$, the simple standard asymptotics
 $\erfc(\sqrt{x})\sim \frac{1}{\sqrt{\pi x}}e^{-x}$ 
 is viable and, when substituted into  \eqref{mainAsymZeroBis:eq}, yields  \eqref{mainAsymptOld:eq}: 
  \begin{align}
    \label{mainAsymptOldBis:eq}
    &\quad\Pbin \notag\\ 
  &\sim 
                 \frac{1}{2}\left[\sqrt{\frac{\gamma}{1-\delta}}  \frac{\sqrt{2}}{\sqrt{\pi}\sqrt{\gamma_s}\sqrt{1-\gamma}}  
    - \frac{1}{\sqrt{\pi \gamma \gamma_s}} \right] 
    \exp\left( {-\frac{\gamma_s(1-\gamma)}{2}}\right) \notag   \\
         &= \frac{1}{2}\left[ \frac{\sqrt{2}\gamma}{\sqrt{1-\delta}\sqrt{1-\gamma}}  
           - 1 \right] \frac{1}{\sqrt{\pi \gamma \gamma_s}} 
           \exp\left( {-\frac{\gamma_s(1-\gamma)}{2}}\right)  \notag \\
         &=
                 \frac{1}{2} \frac{\sqrt{1+\gamma}}{\sqrt{1-\gamma}}  
                  \frac{1}{\sqrt{\pi \gamma \gamma_s}} 
                  \exp\left( {-\frac{\gamma_s(1-\gamma)}{2}}\right) ,
\end{align}
where the last equality can be seen by using 
\begin{equation}
  \sqrt{1+\gamma} + \sqrt{1-\gamma} = \sqrt{2}\sqrt{1+\delta},
\end{equation}
which itself is evident  (from $\delta=\sqrt{1-\gamma^2}$) after squaring.

It is worth recording (cf. \eqref{eq:asymptoticBehavior})  
 that \eqref{mainAsymptOldBis:eq} 
 can be  also rewritten as
  \begin{align}
   \boxed{ \Peb^{m' | m}  
    \sim \frac{1}{\sqrt{2\pi}}
    \frac{   \sqrt{1+\frac{1}{1-\dD^2/2}} }{ \dD } \frac{1}{\sqrt{\gamma_s}}  
    \exp\left( -\frac{1}{2}\gamma_s\frac{\dD^2}{2} \right)
    } 
  \end{align}
  where $\dD$ stands for $\dD(s_m,s_{m'})=\sqrt{2}\sqrt{1-\gamma}$, the incoherent distance 
  between the two symbols 
  $s_m$ and $s_{m'}$ (as given by \eqref{eq:ddDistUnit}).






\nomenclature[$bs$]{$B_s$}{Optical Bandwidth}
\nomenclature[$eta$]{$\eta$}{Spectral efficiency, $\eta:={R_b}/{(N B_s)} = m N^{-1}$}
\nomenclature[$k$]{$k$}{Number of bits per symbol, $k=\log_2(M)$}
\nomenclature[$m$]{$M$}{Constellation cardinality for $M$-ary transmission}
\nomenclature[$rb$]{$R_b$}{Bit rate, $R_b= T_b^{-1}$}
\nomenclature[$rs$]{$R_s$}{Symbol Rate, $R_s = T_s^{-1}$}
\nomenclature[$sigma$]{$\sigma^2$}{Noise variance per quadrature}
\nomenclature[$tb$]{$T_b$}{Bit period}
\nomenclature[$ts$]{$T_s$}{Symbol period}
\nomenclature[$gammab$]{$\gamma_b$}{Bit SNR, $\gamma_b = \gamma_s/k$}
\nomenclature[$gammas$]{$\gamma_s$}{Symbol SNR, $\gamma_s = 1/(2 \sigma^2)$}

\nomenclature[$q1ab$]{$Q_1(a,b)$}{Marcum-Q function of the first kind}
\nomenclature[$i0x$]{$I_0(x)$}{Modified Bessel function}

\nomenclature[$gamma$]{$\gamma$}{Correlation coefficient modulus $\vert \braket{s}{s'} \vert$}

\nomenclature[$pebave$]{$P_{e|b}$}{Bit error probability}

\nomenclature[$pbin$]{$\Pbin$}{Pairwise symbol error probability between $\ket{s_m}$ and $\ket{s_{m'}}$}

\renewcommand{\nomname}{List of frequently-used Symbols}
\setlength{\nomitemsep}{-\parskip}


\section*{Acknowledgments}
This work was supported by the National Science Foundation under Grant 1911183.

The authors would like to thank Prof. M. Karlsson of Chalmers University for bringing to their attention the
decision criterion proposed by \cite{benedetto1994multilevel, Visintin2014}.

\footnotesize
\bibliographystyle{ieeetr}
\bibliography{IEEEabrv,MVMRefs,LatestPapersOpticalInterconnects,LatestMVMarticles}




\end{document}